\documentclass{aa}      

\usepackage{graphicx}
\usepackage{txfonts}
\usepackage{multirow}
\usepackage{natbib}
\usepackage{placeins}
\usepackage[bookmarks=false, colorlinks=true, citecolor=blue, linkcolor=blue]{hyperref}
\usepackage[normalem]{ulem}
\def\kms{\,km\,s$^{-1}$}

\usepackage{color}

\begin{document}

\title{Discovery of recurrent flares of 6.7\,GHz methanol maser emission in  Cepheus\,A\,HW2}

   \author{M. Durjasz
          \inst{1} \href{https://orcid.org/0000-0001-7952-0305}{\includegraphics[scale=0.5]{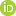}}
          \and
          M. Szymczak
          \inst{1} \href{https://orcid.org/0000-0002-1482-8189}{\includegraphics[scale=0.5]{orcid.png}}
          \and
          M. Olech
          \inst{2} \href{https://orcid.org/0000-0002-0324-7661}{\includegraphics[scale=0.5]{orcid.png}}
          \and
          A. Bartkiewicz
          \inst{1} \href{https://orcid.org/0000-0002-6466-117X}{\includegraphics[scale=0.5]{orcid.png}}
          }

   \institute{Institute of Astronomy, Faculty of Physics, Astronomy and Informatics, Nicolaus Copernicus University, Grudziadzka 5, 87-100 Torun, Poland
   \and
   Space Radio-Diagnostic Research Center, Faculty of Geoengineering, University of Warmia and Mazury\\    Oczapowskiego 2, PL-10-719 Olsztyn, Poland}

  \date{Received XXX / Accepted XXX}

 
  \abstract
{Class II methanol masers at 6.7\,GHz originate close to high-mass young stellar objects (HMYSOs). Their high sensitivity to local condition variations makes them a 
useful marker of the activity of the emerging massive stars. }
{We aim to closely examine the variability of the 6.7\,GHz CH$_3$OH masers in Cep\,A\,HW2 using the new and archival single-dish and high-resolution Very-Long-Baseline Interferometry (VLBI) datasets. }
{We monitored 6.7\,GHz methanol masers towards the target between 2009 and 2021 using the Torun 32\,m radio telescope, and analysed nine datasets of VLBI observations taken over 16\,yr. 
}
{
Faint, extremely redshifted maser emission located close to the presumed central star position and coincident with the radio jet shows 
flaring activity with a period of
$\sim$\enspace5\,yr. Flares have an asymmetric profile with a rise-to-decay time  ratio of 0.18 and relative amplitude higher than 10. The velocity of the flaring cloudlets drifts at a rate of 3-4\,$\times$\,10$^{-5}$\kms\,d$^{-1}$ for about 11.5\,yr of the monitoring. The time-lag between the peaks of the two flaring features implies a propagation speed of the exciting factor of $\sim$\enspace1000\kms. Synchronised and anticorrelated variations of the flux density of blue- and redshifted features begin $\sim$\enspace2.5\,yr after flares of the extremely redshifted emission.
}
{
Our observations suggest that the recurrent flares are the response of low-gain unsaturated maser regions to a relatively low increase in luminosity, which has little effect on the behaviour of most maser regions of higher gain.
}

\titlerunning{Discovery of quasi-periodic, 6.7 GHz methanol maser flare in Cep A }
\authorrunning{M. Durjasz et al.}

\maketitle

\section{Introduction}
The 6.7\,GHz class II methanol maser is the second strongest (after 22\,GHz H$_2$O maser) interstellar emission line and originates from inside the vicinity of high-mass young stellar objects (HMYSOs) \citep{menten1991}. About one-third of these emission lines are associated with ultra-compact HII regions (UCHII) and it is believed that they originate primarily from younger objects, where ionisation processes have not yet developed \citep{hu2016}. Theoretical models show that maser emission originates in regions with number densities up to 10$^9$\,cm$^{-3}$ and dust temperatures higher than 100\,K \citep{sobolev1997,cragg2005}. There is theoretical \citep{sobolev1997} and observational \citep{olech2019, olech2020}  evidence that far-infrared (FIR) photons pump class II CH$_3$OH masers; their relation with IR radiation makes them a good marker of protostellar activity. 

There are currently 1302 class II CH$_3$OH maser sources reported\footnote{\url{https://maserdb.net}} \citep{maserdb_paper}; most are associated with high- and intermediate-mass protostars \citep{szymczak2000, pandian2007,green2010, breen2015, hu2016}. Some of them present substantial variability (e.g. \citealt{szymczak2018, durjasz2019, olech2019, olech2020}), suggesting that protostellar activity fluctuates on time-scales ranging from less than a month up to a few years and more. There are also examples proving that massive protostars also show episodic outbursts that can change maser flux density by more than two orders of magnitude, namely S255$-$NIRS3 \citep{moscadelli2017, szymczak2018}, NGC6334I-MM1B \citep{hunter2018}, and  G358.93$-$0.03 \citep{burns2020}. These phenomena are believed to be caused by an increase in protostar accretion rate which increases star luminosity, effectively boosting the maser pump rate. Although these phenomena result in different flare profiles, they generally show a rapid increase in luminosity followed by much more gradual decay. These flare profiles suggest that maser emission is pumped during an outburst by radiative propagation rather than the physical motion of denser and/or hotter matter. Recent studies revealed thermal propagation in the protostellar disc plane with subluminal ($\geq$ 4\% of the speed of light) velocity \citep{burns2020}. This phenomenon is explained by introducing a {heatwave} energy transfer due to photons, which are absorbed and re-emitted by dust grains.

The last two decades also brought the discovery of the periodic masers \citep{goedhart2003, goedhart2004, goedhart2009, szymczak2011, fujisawa2014, maswanganye2016, szymczak_2016}. Amongst them, there are short-period ($\sim$\enspace1\,month) and long-period (up to a few hundred days) sources; in some cases, only a few spectral features show periodic variations \citep{olech2019}. As the CH$_3$OH masers have been monitored on a regular basis  for over a decade now, we expect that masers with more extended periods will start to be reported, as in the emission  originating from around Cep\,A\,HW2 described in this work. 

Cepheus\,A is a high-mass star-forming region located at a trigonometric distance of 700$\pm$40\,pc \citep{moscadelli2009} that hosts a cluster of YSOs. The brightest continuum source in this cluster is Cep\,A\,HW2 \citep{hw_paper}, which is a HMYSO with a mass of $\sim$\enspace10\,M$_{\odot}$ and bolometric luminosity of  2$\times10^4\,L_{\odot}$ \citep{sanna2017}. \citet{patel2005} revealed a dust emission core with a radius of $\sim$\enspace330\,AU, inclination angle of 26\degr and mass of $\sim$\enspace1\,M$_{\odot}$. HW2 also shows complex outflow activity; \citet{rodrigues1994} estimated the ionised mass-loss rate to about 8$\times$10$^{\mathrm{-7}}$\,M$_{\odot}$\,yr$^{-1}$, and \citet{curiel2006} revealed proper motions of the jet components with an estimated velocity of $\sim$\enspace500\kms. Jets also appeared to be slightly anti-parallel, which could be a result of the precession of the HW2 system. Observations of 2.12\,$\mu$m H$_2$ and 115\,GHz $^{\mathrm{12}}$CO (J=1$-$0)  \citep{cunningham2009} provided further data supporting the hypothesis of precession, presumably driven by a companion in an eccentric, non-coplanar orbit that triggers pulsing activity with passage through its periastron every $\sim$\enspace2500\,yr. 

Observations of the 22$\,$GHz water vapour maser  \citep{torrelles_1996, torrelles_2001} revealed complex morphology with arc-like structures, which  appeared to be tracing slow shock waves propagating through a rotating disc \citep{gallimore_2003}. Multi-epoch H$_2$O maser imaging \citep{torrelles_2011} revealed the simultaneous presence of a wide-angle (102\degr\,), slow (13 - 18\kms) outflow and narrow-angle (18\degr) ionised, fast (70\kms) jet. 

Recent 44\,GHz continuum imaging \citep{carrasco2021} revealed that slow outflow originates from a region at a distance of $\sim$\enspace3\,AU from the HW2 protostar and is collimated into a jet $\sim$\enspace25\,AU away. The proposed scenario of external collimation is the presence of a large-scale magnetic field and a dense ambient medium. \citet{vlemmings2010} reported the presence of a large-scale magnetic field of $\sim$\enspace23\,mG directed perpendicularly to the disc plane, suggesting that in this case the accretion onto the disc plane is likely regulated by magnetic forces.

 \citet{menten1991} were the first to report 6.7\,GHz CH$_3$OH masers in Cep\,A\,HW2 with a peak flux density F$_{\mathrm{6.7}}$\,=\,1420\,Jy at V$_{\mathrm{lsr}}\,\simeq$\,2.5\kms. Subsequent observations did not show significant variations in the profile shape but substantial variations in the amplitude, which reached a factor of 3 over $\sim$\enspace25\,yr \citep{szymczak2000, vlemmings_2008, sugiyama_2008, fontani_2010, szymczak2012, hu2016, sanna2017, yang_2017}. Synchronised and anticorrelated variations between the three most blueshifted and two most redshifted spectral features were reported by \citet{sugiyama_2008}. VLBI observations of the methanol masers revealed that they originate from a distance of between 300 and 1000\,AU from the presumed position of HW2 and present arc-like structure with velocity field showing no signs of rotation, but consistent with the infall scenario \citep{torstensson_2011}. 
Proper-motion measurements \citep{sugiyama_2014, sanna2017} confirmed this scheme, which suggests that disc accretion is the most likely scenario of the formation of Cep\,A\,HW2. Long-term monitoring \citep{szymczak2014} also revealed the episodic presence of the redshifted low-amplitude spectral features and radial velocity drifts of these and a few persistent features. 

In the present work, we extend the \citet{szymczak2014} dataset by reporting results from 12 years of single-dish monitoring and VLBI imaging of the CH$_3$OH masers in Cep\,A\,HW2. Our main result is  the detection of periodic ($\simeq$ 5\,yr) variability of the  redshifted faint emission.

\section{Observations}
\subsection{Single-dish observations} \label{sec:single-dish}
We used new and archival \citep{szymczak2014} 6.7\,GHz data obtained with the Torun 32\,m radio telescope.
We observed the target source as part of a monitoring program from June 2009 to December 2021 and acquired new data from March 2013 with a typical cadence of nine observations per month. There were several gaps of 3-4 weeks. Due to telescope maintenance, we took no single-dish spectra between 2020 May and October. 

The full-beam width at half maximum of the antenna at 6.7\,GHz was 5\farcm8, and the pointing error was $\sim$\enspace25\arcsec\, before mid-2016 and $\sim$\enspace10\arcsec\, later \citep{lew2018}. The system temperature ranged from 25 to 40\,K. The data were dual-polarisation taken in frequency switching mode. We used the autocorrelation spectrometer to acquire spectra with a resolution of 0.09\kms\, after Hanning smoothing and a typical 1$\sigma$ noise level of 0.35\,Jy before May 2015 and 0.25\,Jy afterward. We based the flux density scale on continuum observations of 3C123 and spectra of little variable methanol maser source G32.744$-$0.076 \citep{szymczak2014}. The resulting accuracy of the absolute flux density was better than 10\%.
\subsection{European VLBI Network observations}
We carried out European VLBI Network (EVN)\footnote{The EVN is a joint facility of European, Chinese, South African and other radio astronomy institutes funded by their national research councils.} observations in June and October 2020;  project codes were RD002 and ED048B, respectively (Table\,\ref{tab:evn_projs}). Phase-referencing observations of the 6.7\,GHz methanol maser line were performed and correlated with two polarisation combinations (RR, LL). J2302+6405 was used as a phase-referencing calibrator and 2007+777 as a fringe finder. J2254+6209 was used as a phase-referencing calibrator for the RD002 project. Scans on the target and phase calibrators were performed in 5\,min cycle: 3\,min 15\,s on Cep\,A, 1\,min 45\,s on phase calibrator. The data were processed with the SFXC software correlator \citep{keimpema2015} at the Joint Institute for VLBI in Europe using 1\,s averaging time and two frequency setups. The first setup for high spectral sampling used 2048 channels over a 4\,MHz band (2\,MHz for RD002), yielding channel spacing of 0.088\kms\, (0.044 for RD002) and the second setup used 8\,$\times$\,4\,MHz band with 128 channels to improve the sensitivity of the calibrator maps. 
We reduced the data using the NRAO Astronomical Image Processing System (AIPS), following standard procedures described in the EVN data reduction guide\footnote{\url{https://www.evlbi.org/evn-data-reduction-guide}}. Fringe fitting in both projects was done using phase-referencing calibrators to improve astrometry. We produced intensity maps (Stokes I) within an area of $2\arcsec\times2\arcsec\,$ around the phase-centre and estimated the intensity of maser emission by fitting a 2D Gaussian function (task JMFIT in NRAO AIPS package). The final 1$\sigma$ noise level in both projects was better than 6\,mJy\,beam$^{-1}$ for emission-free channels.

\subsection{Archival VLBI data}
In this work, we also make use of the publicly available EVN \citep{torstensson_2011, sanna2017} data and Japanese VLBI Network (JVN; \citealt{sugiyama_2014}) observational results for a total of seven epochs. Archival EVN data from experiments ES071A, ES071B, and ES071C \citep{sanna2017} were reduced, while for the rest, the parameters of maser spots were taken from published tables (\citealt{torstensson_2011,sugiyama_2014}, their Tables 3 and 5, respectively). Details of all VLBI projects used in this paper are listed in Table \ref{tab:evn_projs}.

\begin{table*}
\caption{VLBI observations used in the paper. \label{tab:evn_projs} }
\centering
\begin{tabular}{lcccccccc}
\hline
 & Epoch & &  Exp. & Telescope$^a$  & Vel. res. & 1$\sigma$ noise & Beam & Ref. \\\cline{1-3}
No. & Date & MJD       &  code &          & (\kms)       & (mJy\,beam$^{-1}$) & (mas x mas, \degr) &  \\
\hline
1 & 2004 Nov. 06 & 53315 & EL032  & EVN(8) & 0.09  & 7     & 13.8$\times$5.3,  $-$57 & 1 \\
2 & 2006 Sep. 09 & 53987 &        & JVN(4) & 0.18  & 60    &  9.4$\times$4.3,  $-$70 & 2 \\
3 & 2007 Jul. 28 & 54309 &        & JVN(5) & 0.18  & 90    &  9.2$\times$4.3,  $-$43 & 2 \\
4 & 2008 Oct. 25 & 54764 &        & JVN(5) & 0.18  & 120   &  7.0$\times$3.9,  $-$79 & 2 \\
5 & 2013 Mar. 01 & 56352 & ES071A & EVN(8) & 0.04  & 4$-$6 &  4.3$\times$3.1,  $-$58 & 3 \\
6 & 2014 Feb. 28 & 56716 & ES071B & EVN(8) & 0.04  & 4$-$6 &  4.0$\times$3.4,  $-$67 & 3 \\
7 & 2015 Mar. 13 & 57094 & ES071C & EVN(9) & 0.04  & 4$-$6 &  3.9$\times$3.0,  $-$53 & 3 \\
8 & 2020 Jun. 02 & 59002 & RD002  & EVN(4) & 0.04  & 5     &  7.5$\times$3.5,    +86 & 4 \\
9 & 2020 Oct. 18 & 59140 & ED048B & EVN(8) & 0.09  & 4     &  4.5$\times$3.8,    +12 & 4 \\
\hline
\end{tabular}
\tablebib{
(1)~\citet{torstensson_2011}; (2) \citet{sugiyama_2014}; (3) \citet{sanna2017}; (4) this work.
}
\tablefoot{$^a$ 
Telescopes are the EVN and the Japan VLBI Network (JVN). The number of antennas is given in parentheses.
}
\end{table*}

\section{Results}
\begin{figure}
\centering
\includegraphics[width=1\columnwidth]{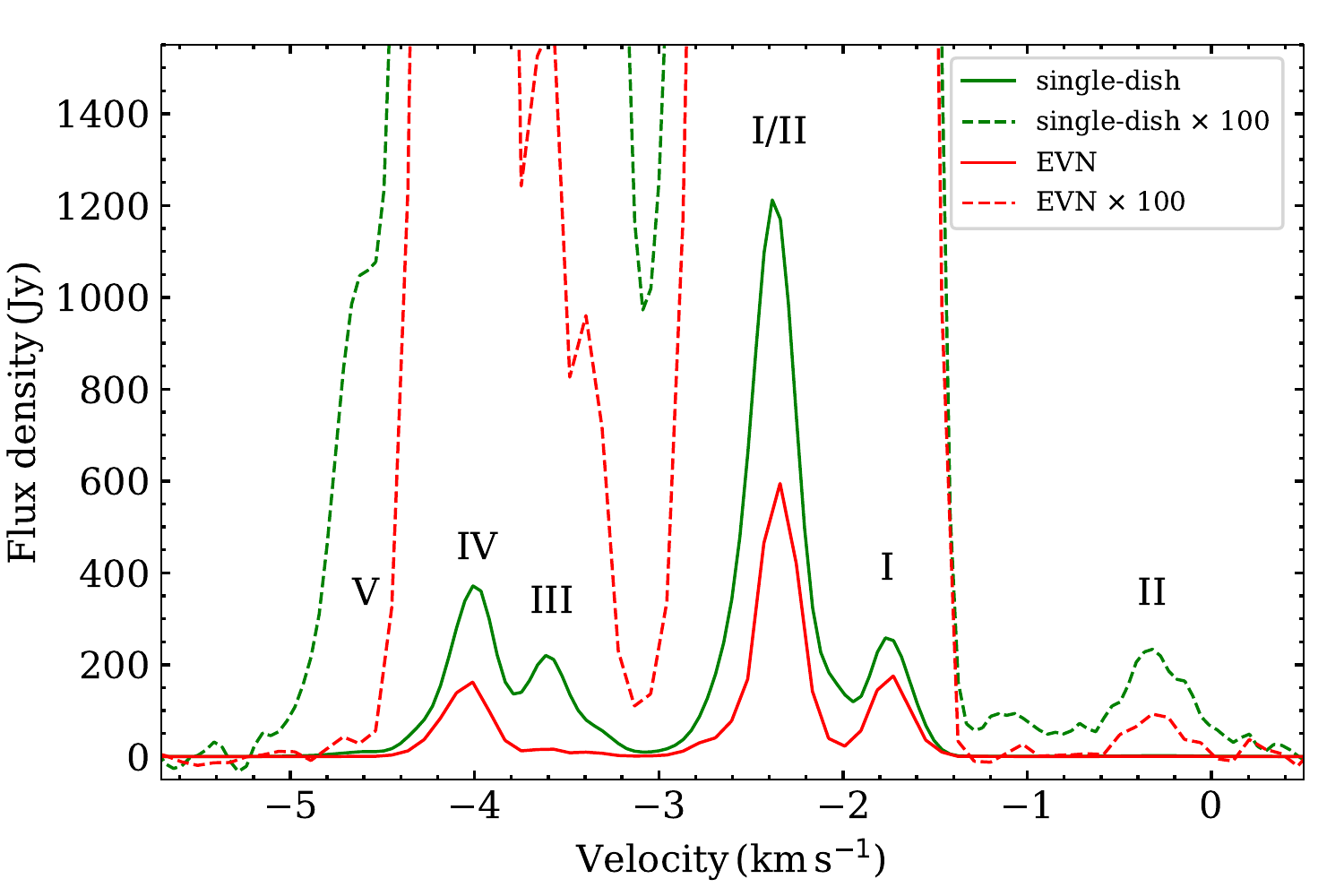}
\caption{6.7\,GHz maser single-dish auto-correlation spectrum of Cep\,A\,HW2 ({\it {green}}) obtained on 2020 October 18 and the cross-correlated spectrum from EVN observations ({\it {red}}) taken on the same date. The dashed lines represent the same spectra magnified 100 times to show low-flux-density features. The features are marked by roman numerals following \citet{sugiyama_2008}, which correspond to clusters shown in Fig.\,\ref{fig:cepa-spots-june-and-october-2020}. \label{fig:cepa_spec_last} }
\end{figure}

\begin{figure*}
\centering
\includegraphics[width=0.9\textwidth]{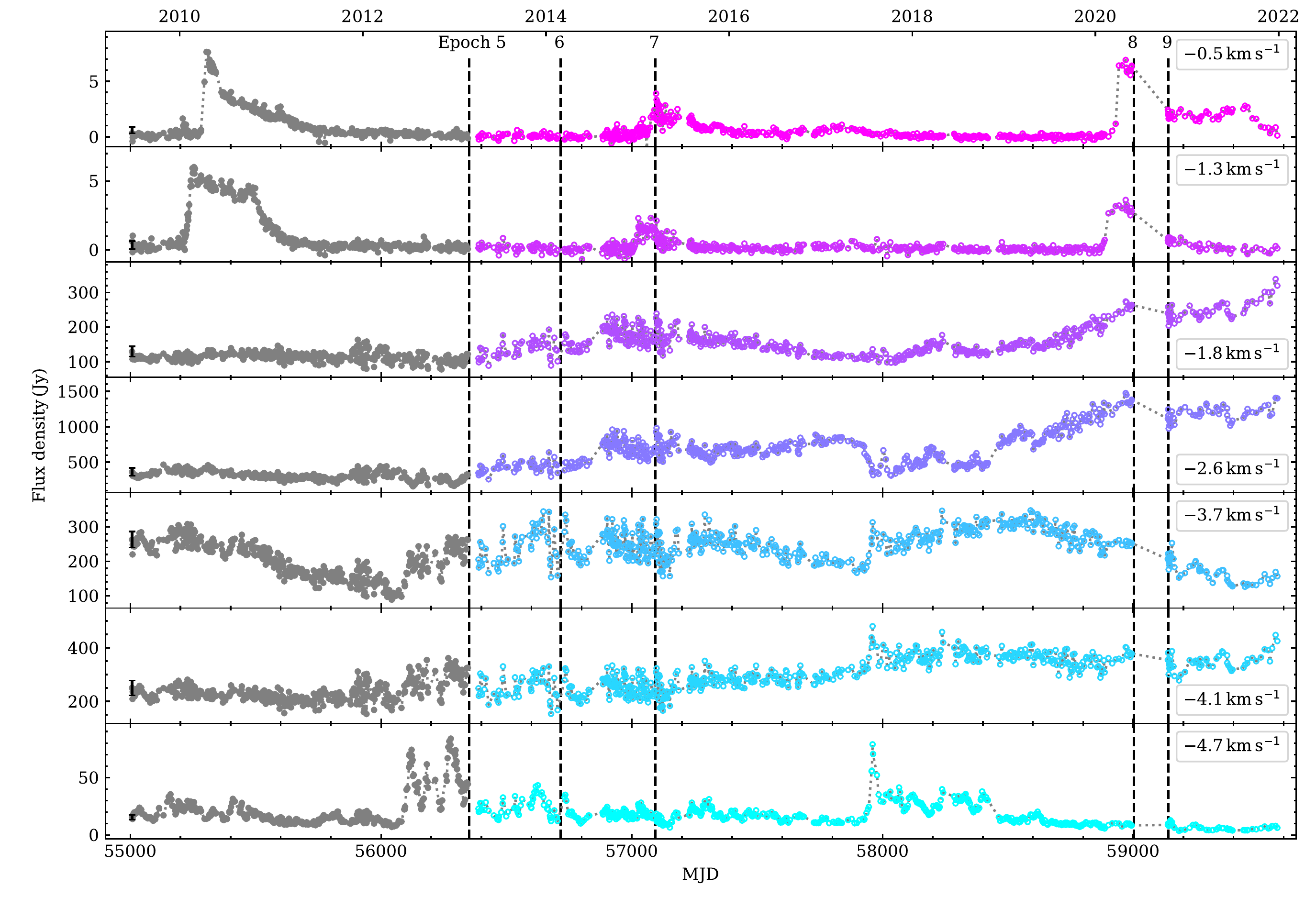}

\caption{Light curves of the main spectral features of the 6.7\,GHz transition in Cep\,A\,HW2. Data before MJD 56352 (grey dots) were published in \citet{szymczak2014}. Typical measurement uncertainty is shown by the black bar for the first data point. The light curves were produced using single-dish data (Sect.\,\ref{sec:single-dish}) by averaging three channels; the velocity of the central channel is given for each panel. Colours indicate different spectral features at velocities given to the right of the plots.
The vertical black dashed lines mark VLBI observations; epoch numbers correspond to those given in Table\,\ref{tab:evn_projs}. VLBI observations at epochs 1-4 were performed before our single-dish monitoring started. \label{fig:cepa_lcs} } 

\end{figure*}

We find that in addition to five persistent features known since discovery \citep{menten1991}, the 6.7\,GHz maser spectrum of the target has two intermittent low-intensity features (Fig.\,\ref{fig:cepa_spec_last}). At the beginning of the first flare observed during the monitoring, these two most redshifted features peaked at $-$1.3 and $-$0.5\kms. This designation is used hereafter for clarity, even though both features drift in velocity (Sect.\,\ref{sec:flares_description}). Light curves of the main features (Fig.\,\ref{fig:cepa_lcs}) illustrate significant variability of the emission. They present previously reported synchronous anticorrelation of red- and blueshifted features \citep{sugiyama_2008, szymczak2014} and the cyclical activity of the most redshifted, weak features over $\sim$\enspace12\,yr are revealed for the first time. 

\begin{figure}
\centering
\includegraphics[width=1\columnwidth]{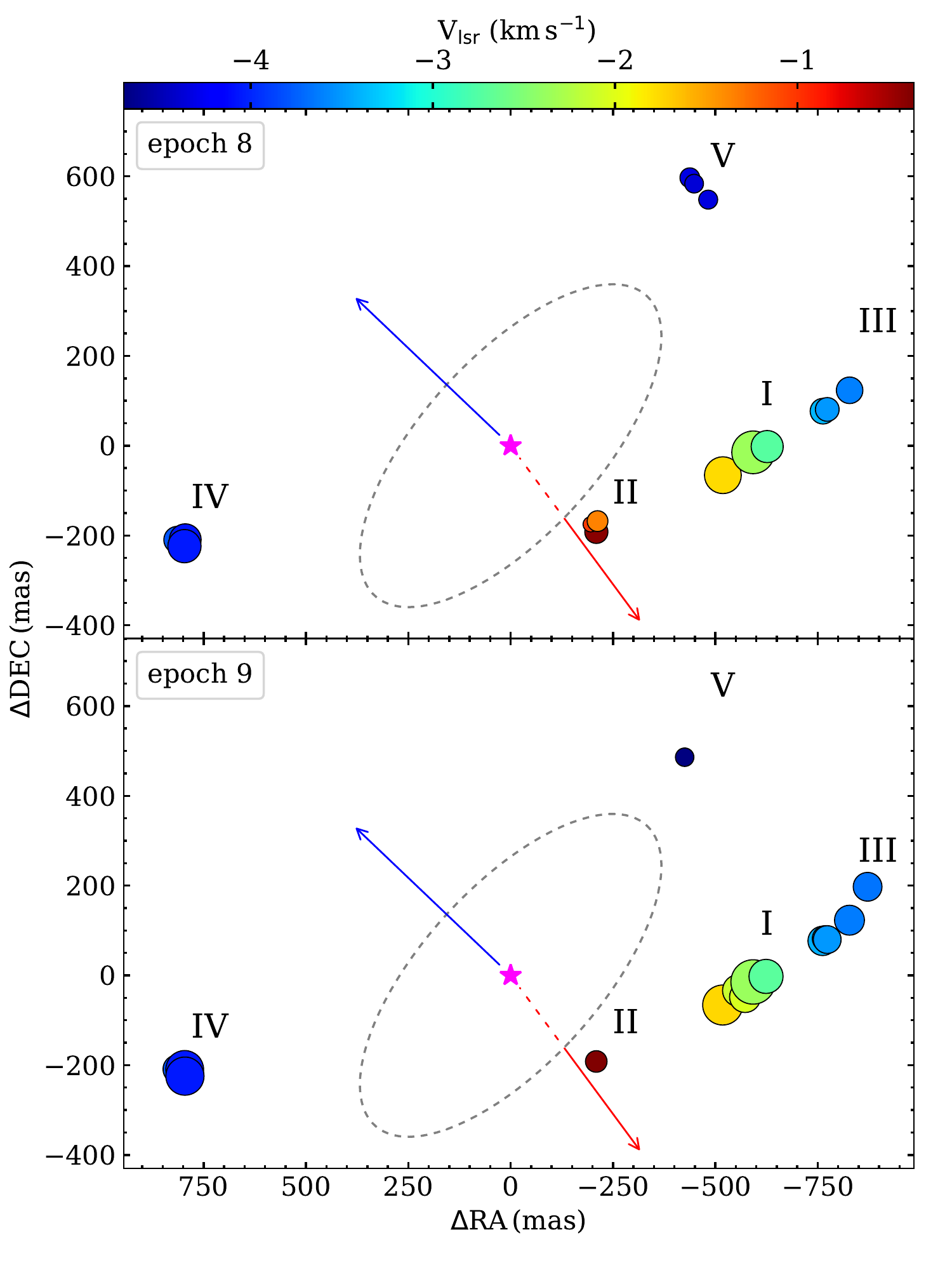}
\caption{Spatial distributions of the 6.7\,GHz methanol maser cloudlets in Cep\,A\,HW2 in epochs 8 and 9. The circle size is scaled as the logarithm of peak brightness; its colour corresponds to the local standard of rest velocity scale shown in the wedge. Roman units indicate the cloudlet numbering according to \citet{sugiyama_2014} and \citet{sanna2017}. The magenta star marks the Cep\,A\,HW2 position \citep{curiel2006}: $\alpha$(J2000)=22$^\mathrm{h}$56$^\mathrm{m}$17.9816$^\mathrm{s}$, $\delta$(J2000)=62\degr01$'$49.572$''$.
The ellipse denotes the dust disc emission at 0.9\,mm \citep{patel2005}, and the arrows mark the elongated knot directions at the base of the jet detected at 7.5\,mm \citep{carrasco2021}.   \label{fig:cepa-spots-june-and-october-2020} }
\end{figure}

In this publication, we adopt the definition of {\it spot} and {\it cloudlet} provided by  \citet{sanna2017}. New VLBI data are summarised in Figure\,\ref{fig:cepa-spots-june-and-october-2020}, showing the spatial distribution of the maser cloudlets, and in Tables\,\ref{tab:spots_rd002} and \ref{tab:spots_ed048B}, which list the spot parameters. We detected 90 and 71 individual maser spots above the 5\,$\sigma$ threshold in epochs 8 and 9, respectively. The main properties of cloudlets derived from our EVN observations in epochs 8 and 9 are given in Table\,\ref{tab:cloudlet_gauss_fit_results}. In addition, the parameters of the corresponding cloudlets detected in epochs 5$-$7 are added. 

In our observation epochs, the maser cloudlets were distributed over five clusters (Figure\,\ref{fig:cepa-spots-june-and-october-2020}), forming the arched structure well known from previous studies \citep{vlemmings2010, torstensson_2011, sugiyama_2014, sanna2017}; this implies the overall 6.7\,GHz maser structure is stable for $\sim$\enspace15\,yr. 

\subsection{Recurrent flares of the most redshifted emission} \label{sec:flares_description}
The features $-$1.3 and $-$0.5\kms\, were detected three times during our 12-year monitoring period. The parameters of the flares are listed in Table\,\ref{tab:flares}, where we assume the start and end of burst as the nearest data points with flux density below $3\sigma$ before and after the data point which has flux density exceeding $5\sigma$ (similarly to \citealt{pietka2015}). The intervals between the onset of subsequent flares are 1796, 1870\,d and 1787, 1859\,d for $-$1.3 and $-$0.5\kms\, features, respectively, resulting in an average period of 1828\,d with the range of 168\,d. Flares at $-$1.3 and $-$0.5\kms\, lasted 99$-$387 and 227$-$583\,d, respectively, which are on average 13\% and 22\% of the period. The flared profile is highly asymmetric; the average rise and decay times are 35d and 200\,d ($-$1.3\kms), and 32d and 373\,d ($-$0.5\kms). We notice that the flare parameters for the $-$1.3\kms\ feature show greater dispersion than those for $-$0.5\kms.

\begin{figure}
\centering
\includegraphics[width=0.9\columnwidth]{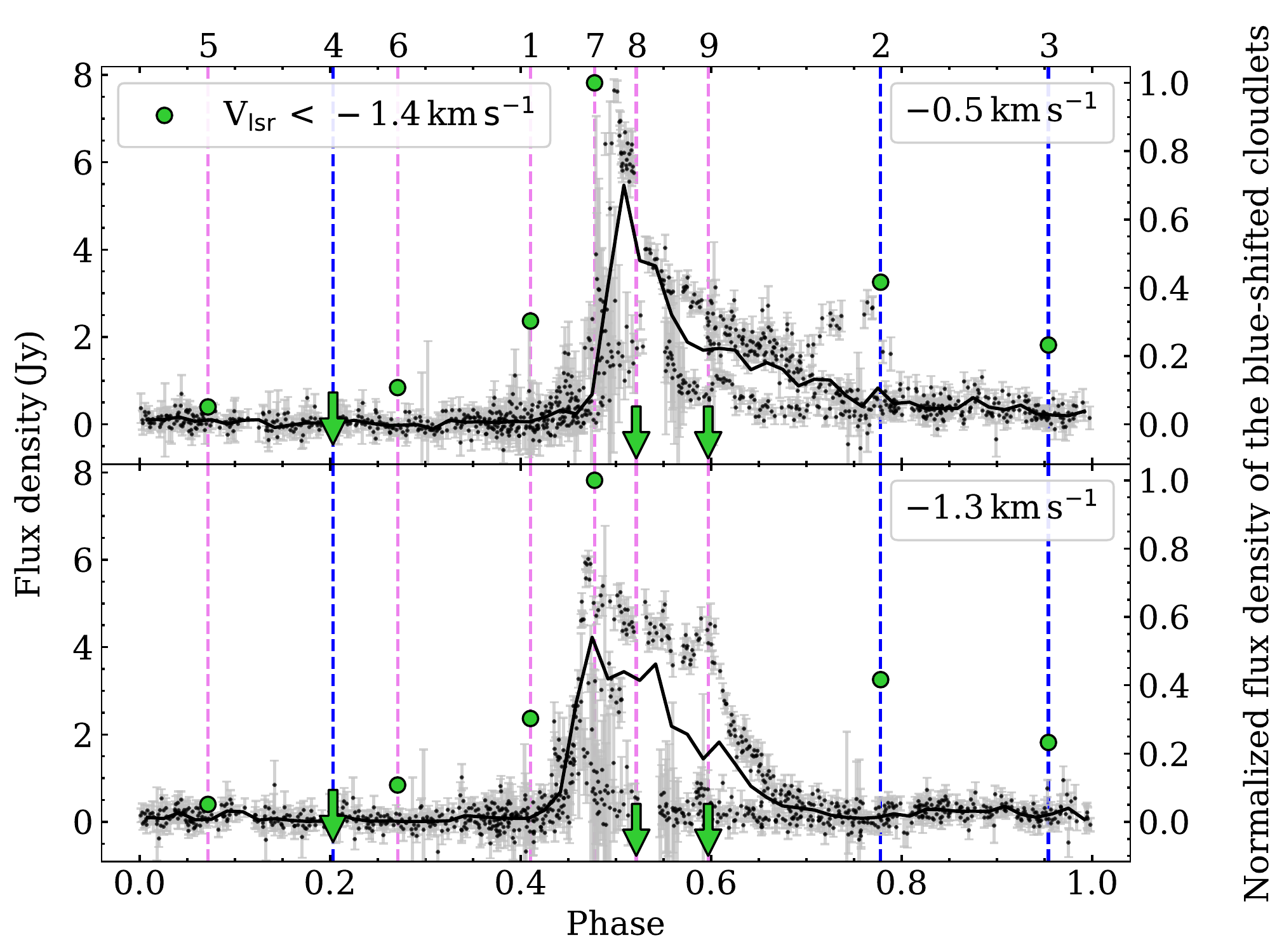}
\caption{Folded light curves of the redshifted flaring emission for $-$0.5 (top) and $-$1.3\kms\,(bottom) features. The data points are black with grey error bars, while the black solid lines show the average profiles. Green points refer to the relative intensities of the blueshifted emission observable in cluster II; arrows mark upper limits for the blueshifted emission in epochs with no detection. The vertical dashed lines correspond to the computed phases of VLBI observations (magenta - EVN, blue - JVN); the epoch numbers (Table\,\ref{tab:evn_projs}) are given above the top abscissa.
\label{fig:folded_light_curve} }
\end{figure}

Figure\,\ref{fig:folded_light_curve} shows the time series of the two redshifted features folded with a period of 1828\,d. The flare duration and peak flux density differ significantly from cycle to cycle. At $-$1.3\kms\, , the flare onset occurs 41--61\,d earlier than that at $-$0.5\kms.

\begin{figure}
\centering
\includegraphics[width=0.9\columnwidth]{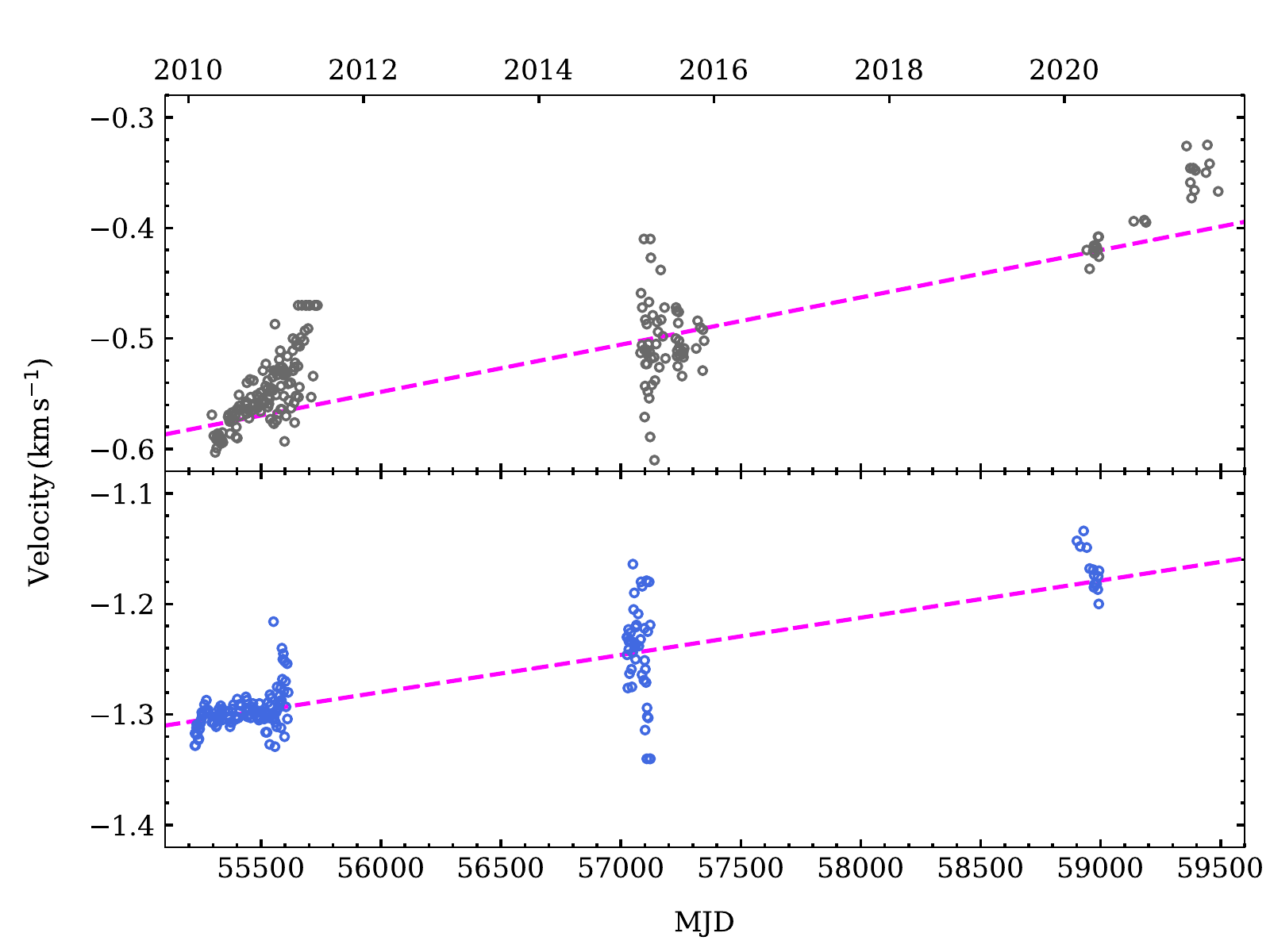}
\caption{Drifts of radial velocity of the $-$0.5 (top) and $-$1.3\,\kms\, (bottom) features. The magenta dashed line marks the linear function fit. \label{fig:radial_curve}}
\end{figure}

During the first flare, both flaring features exhibited a systematic drift of peak velocity \citep{szymczak2014}. The present study extends the data for two subsequent cycles. To quantify the rate of velocity drift, we fitted the Gaussian function to the spectra using the CURVE\_FIT method from the SCIPY.OPTIMIZE package \citep{2020SciPy-NMeth}. The following intervals are analysed: MJD 55224-55792 (flare \#1), 57007-57488 (flare \#2), and 58882-59792 (flare \#3). The result obtained is shown in Figure\,\ref{fig:radial_curve}. The rates of velocity drift are 0.015$\,\pm\,$0.001 and 0.012$\,\pm\,$0.001\,\kms\,yr$^{-1}$ for the features at $-$0.5\,\kms\, and $-$1.3\,\kms\,, respectively. For the feature at $-$0.5\kms\, , there is evidence of an increase in drift rate at the end of flares \#1 and \#3 that deviates from the average value derived from the long-term data. 

\begin{table}
\centering
\caption{Flare parameters of the $-$1.3 and $-$0.5\kms\, features. \label{tab:flares} }
\begin{tabular}[c]{ccccc}
\hline
$V_{\mathrm{lsr}}$ & $t_{\mathrm{start}}$ & $t_{\mathrm{peak}}$ & $t_{\mathrm{end}}$ & $F_{\mathrm{peak}}$ \\
(\kms) & (MJD) & (MJD) & (MJD) & (Jy) \\
\hline
\multicolumn{5}{c}{Flare \#1} \\
\hline
$-$1.3 & 55222 & 55255 & 55609 & 6.0 \\
$-$0.5 & 55283 & 55307 & 55679 & 7.6\\
\hline
\multicolumn{5}{c}{Flare \#2} \\
\hline
$-$1.3 & 57018 & 57025 & 57117 & 2.7 \\
$-$0.5 & 57070 & 57095 & 57297 & 3.3 \\
\hline
\multicolumn{5}{c}{Flare \#3} \\
\hline
$-$1.3 & 58888 & 58953 & 59104 & 3.5 \\
$-$0.5 & 58929 & 58966 & 59512 & 7.0 \\
\hline
\end{tabular}
\tablefoot{$V_\mathrm{lsr}$ is the peak velocity at the beginning of the first flare,  
$t_\mathrm{start}$, $t_\mathrm{peak}$, and $t_\mathrm{end}$, are the start, maximum, and end times of the flare, respectively, and $F_\mathrm{peak}$ is the peak flux density.
}
\end{table}

\subsection{Position and properties of the most redshifted cloudlets}\label{sec:pos-prop-red}
\begin{figure}
\centering
\includegraphics[width=0.45\textwidth]{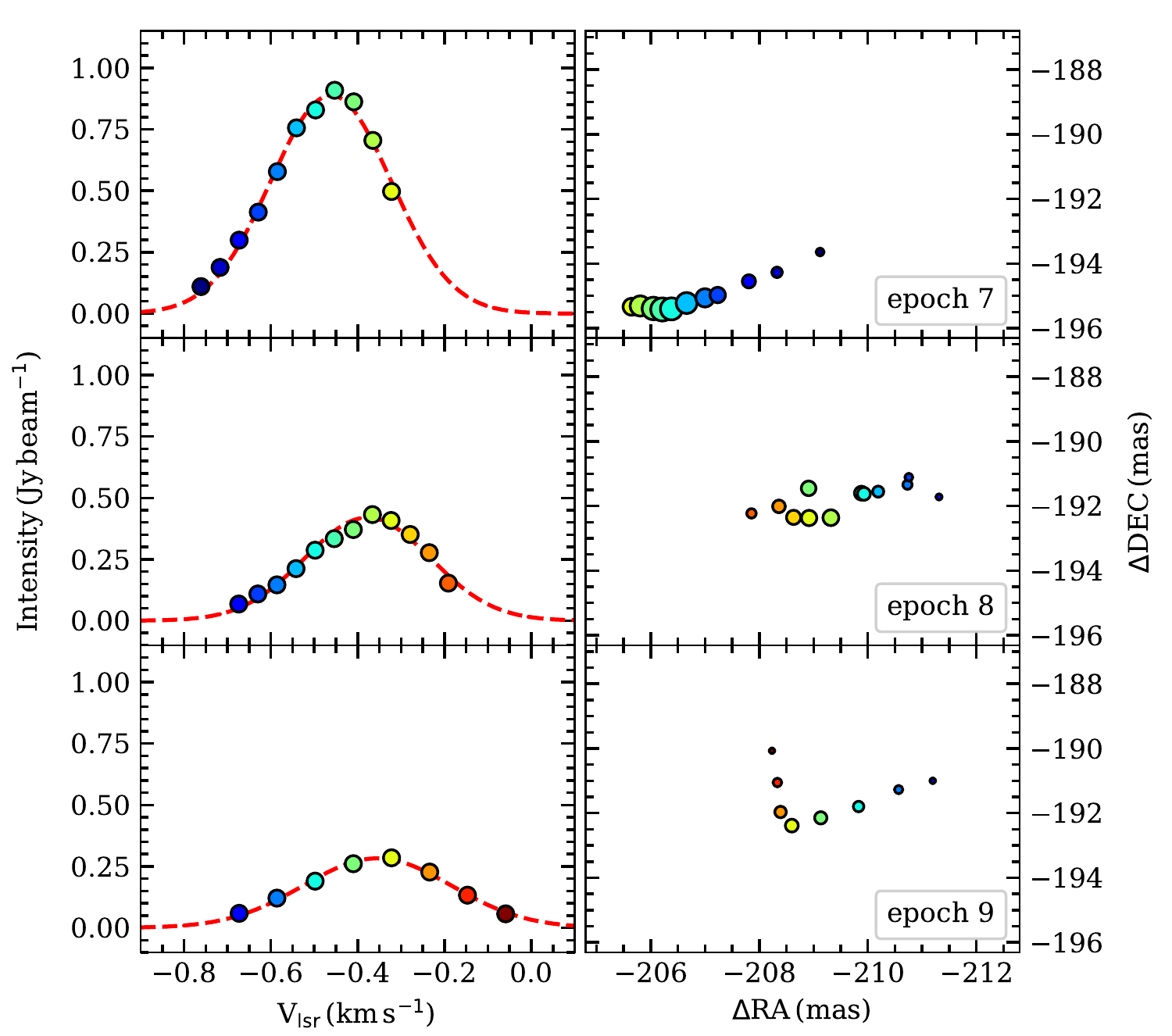}
\includegraphics[width=0.45\textwidth]{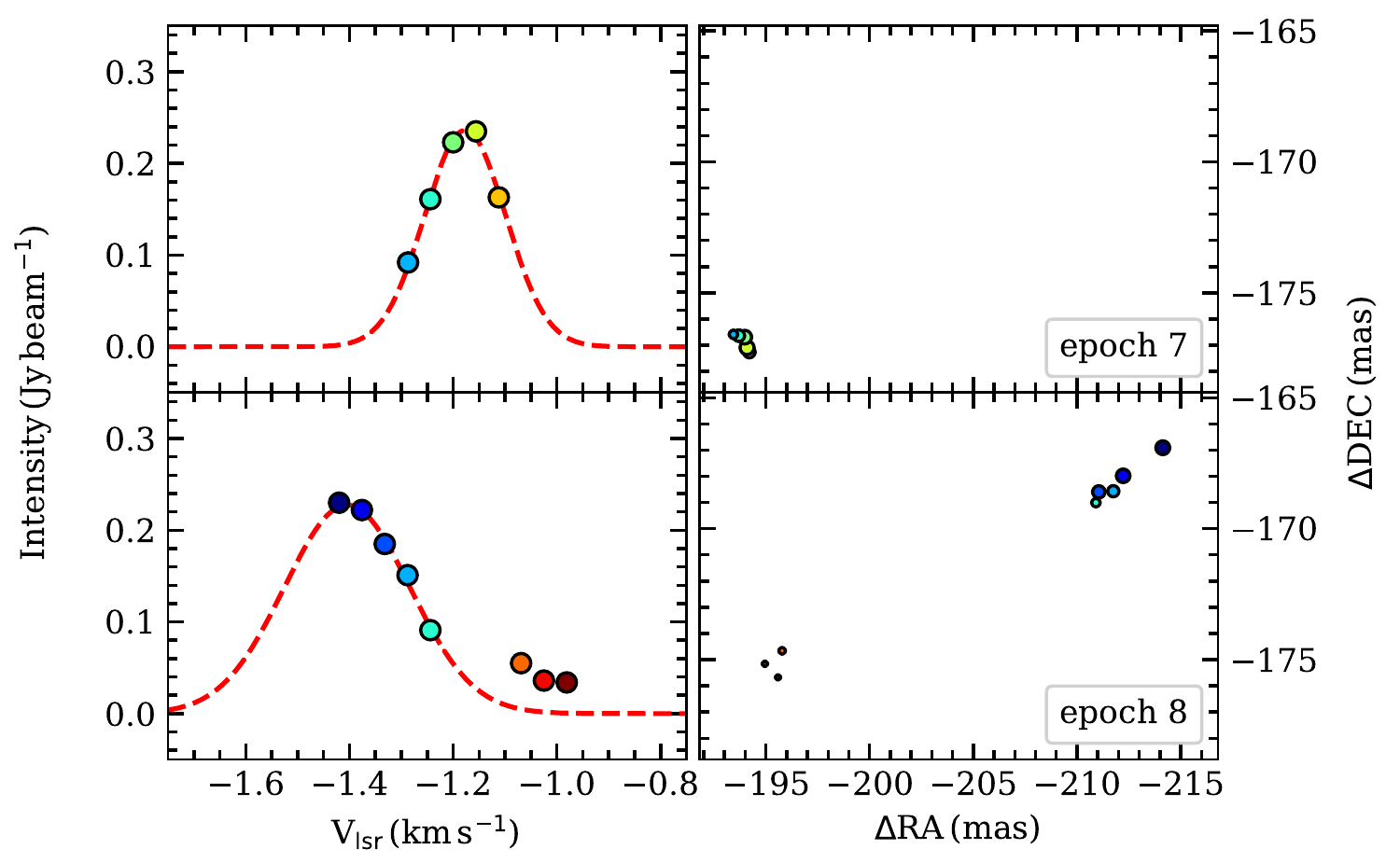}
\caption{Line profiles (left) and distributions (right) of the maser spots in the two most redshifted cloudlets at the indicated epochs (Table\,\ref{tab:evn_projs}). Each symbol represents a methanol maser spot. The red dashed line denotes the Gaussian function fit (Table\,\ref{tab:cloudlet_gauss_fit_results}). The symbol sizes in the right column are proportional to the square root of brightness, whereas the colours refer to velocity. 
} \label{fig:maps_redshifted_flares_with_gauss}
\end{figure}

The emission of the $-$0.5\kms\, feature originates from cluster II (Fig.\,\ref{fig:cepa-spots-june-and-october-2020}, Tables\,\ref{tab:spots_rd002} and \ref{tab:spots_ed048B}). EVN observations in epoch 8 revealed the emission of $-$0.5\kms\, feature with the peak brightness of 0.42\,Jy\,beam$^{-1}$ at $-$0.38\kms\, and with FWHM equal to 0.34\kms. After 4.5\,months, in epoch 9, the brightness decreased to 0.28\,Jy\,beam$^{-1}$, the peak velocity shifted to $-$0.30\kms\, and FWHM increased to 0.40\kms\, (Fig.\,\ref{fig:maps_redshifted_flares_with_gauss}). This feature in the 32 m spectrum taken on 2020 May 24, a week before the EVN experiment, reached a peak flux density of 6.2\,Jy, indicating that about 93\% of the maser flux is missing; thus, this flaring emission is highly resolved out with the EVN beam. 

In epoch 8, the $-$0.5\kms\, cloudlet has a linear structure of about 3\,mas\,$\times$\,2.5\,mas in size with the velocity gradient of 4.19\,mas/\kms\, along with PA\,$\simeq$\,$-$75\degr\, (Fig.\,\ref{fig:maps_redshifted_flares_with_gauss}). In epoch 9, the blue part of the cloudlet is preserved, while the red part folds into a hook. A slightly curved morphology was observed in epoch 7 by \citet{sanna2017} (Fig.\,\ref{fig:maps_redshifted_flares_with_gauss}), but its general appearance resembles that seen in our data. Details of cloudlet parameters from Gaussian fitting are presented in Table\,\ref{tab:cloudlet_gauss_fit_results}.

We note that the peak velocities inferred from EVN observations in epochs 7-9 are consistent with values obtained from the single-dish spectra and confirm the rate of velocity drift of 0.015\kms\,yr$^{-1}$.

No emission at $-$0.5\kms\, was detected in epochs 5 and 6 \citep{sanna2017} when the RMS noise level in a line-free channel was $\sim$\enspace5\,mJy\,beam$^{-1}$, which corresponds to a brightness temperature of 8\,$\times$\,10$^6$\,K. During the flare, in epochs 8 and 9, the brightness temperature of the brightest part of the cloudlet has increased to 2.2\,$\times$\,10$^8$ and 1.5\,$\times$\,10$^8$\,K, respectively.

Using the procedure outlined by   Sanna et al.  and the EVN data from epochs 7 and 9, we estimated that the proper motion of the $-$0.5\kms\, cloudlet is V$_{\mathrm{RA}}$\,=\,$-$2.06 $\pm$ 0.12\kms,  V$_{\mathrm{DEC}}$\,=\,2.91\,$\pm$\,0.17\kms; which implies infall and rotation velocity components of 4.0 and 3.8\kms, respectively. Therefore, the rotational velocity is about 85\% of the expected Keplerian velocity at a distance of 450\,AU for the assumed mass of HW2 of 10\,M$_{\odot}$.

Our EVN data imply that the $-$1.3\kms\, emission was only detected in epoch 8 as two groups of spots separated in position ($>$15\,mas) and velocity (Fig.\,\ref{fig:maps_redshifted_flares_with_gauss}). The profile of the group with the intensity peaking near $-$1.4\kms\,  appears as an incomplete Gaussian, while that peaking near $-$1.1\kms\, is irregular. The latter, composed of three spots in contiguous channels, has an intensity slightly above the 5\,$\sigma$ level and a brightness temperature of 2$\times$10$^{7}$\,K; its position is very close to that observed in epoch 7, and the difference in velocity between the two epochs is consistent with the velocity drift trend inferred from the single-dish observations. A comparison of the cross-correlation and auto-correlation spectra reveals that the $-$1.3\kms\, emission is highly resolved out; the missing flux is about 95\%. We suggest that significant evolution of this redshifted emission took place over a timescale of 5\,yr. However, as the data in epoch 8 were obtained with use of a limited number of antennas and the map dynamic and angular resolution is lower than in the other epochs, the observed morphology and profile may be distorted. 

Both $-$0.5 and $-$1.3\kms\, cloudlets are located close to each other at a distance of 35\,mas, which corresponds to 50\,AU when the inclination angle of 64\degr\, and position angle of 134\degr\, \citep{sanna2017} are taken into account. For the presumed position of the star proposed by \citet{curiel2006}, the corresponding distances are 650\,mas (450\,AU) and 600\,mas (420\,AU).
Assuming that the central source produces the flare, a triggering factor would first reach the $-$1.3\kms\, and then the $-$0.5\kms\, cloudlet; this is in good agreement with the single-dish monitoring results, which revealed time-lags between the onset of the flares. The time-lag of 51\,d estimated for the 2015 flare ($\sim$\enspace$ $MJD 57020) implies that an exciting factor is moving with a velocity of $\sim$\enspace1000\kms\, in the plane of the maser disc.

\subsection{Long-term variability of cluster II}
Figure \ref{fig:cluster_I_and_II_8_epochs} presents the spot maps of cluster II in nine epochs of VLBI observations. The redshifted flaring emission was visible in epochs 2, 7, 8, and 9. These epochs were in active state assuming a variation period of 1828\,d (Fig.\,\ref{fig:folded_light_curve}). Our EVN measurements provide evidence that the flaring emission comes preferentially from cluster II which is located very close to the outer edge of the dust disc \citep{patel2005, sanna2017}. In six VLBI observation epochs, the emission peaking at $-$2.4\kms\, (Fig.\,\ref{fig:cluster_I_and_II_8_epochs}) was detected in cluster II.

\subsection{Synchronous and anticorrelated variability}
\begin{figure}
\centering
\includegraphics[width=0.9\columnwidth]{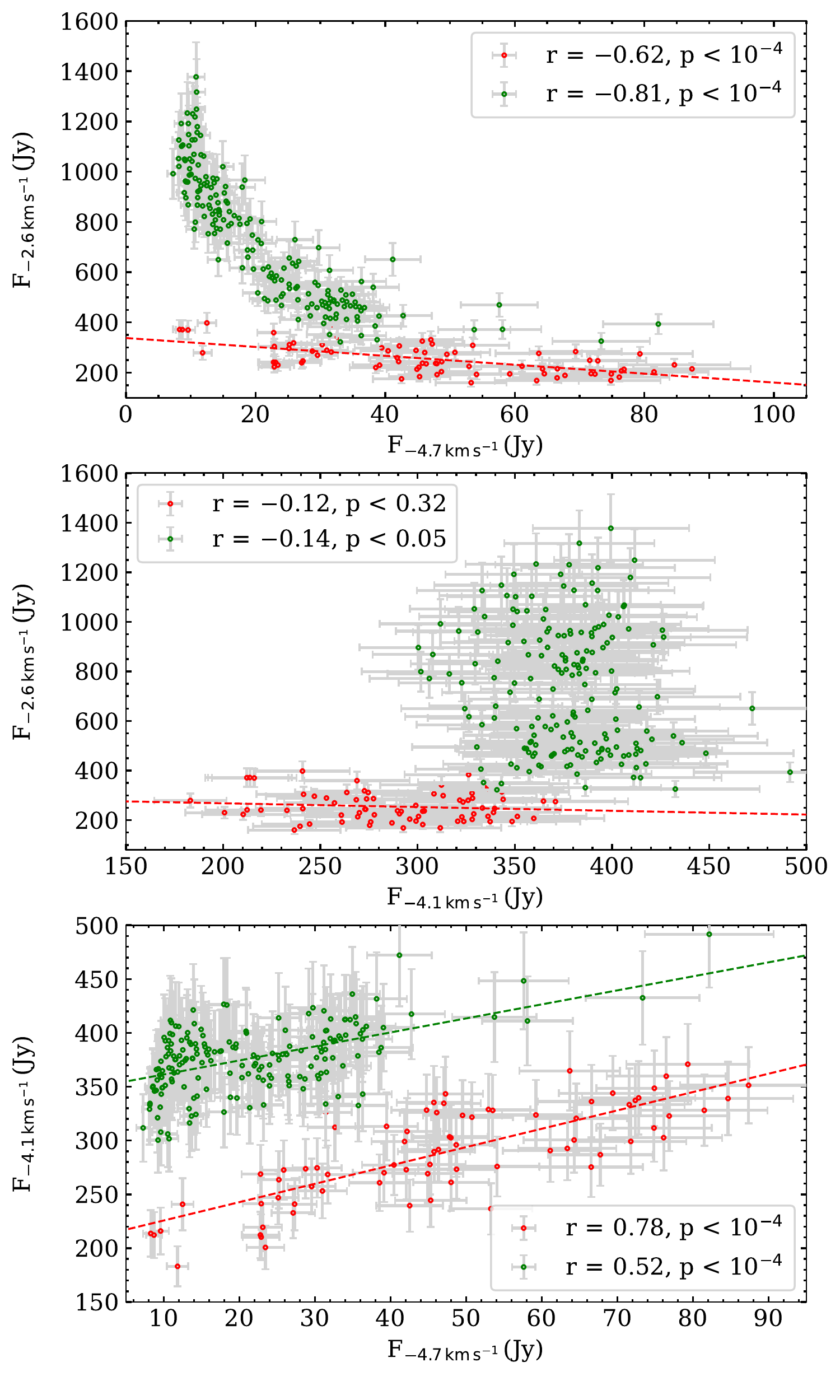}
\caption{Comparison of peak flux densities of the selected features. Red and green points refer to measurements from intervals of MJD\,56057-56347 (similarly as in \citealt{szymczak2014}) and MJD\,57862-58851, respectively. The dashed lines mark the best fit of the linear function.
\label{fig:cepa-anticorr} }
\end{figure}

\begin{figure}
\centering
\includegraphics[width=0.9\columnwidth]{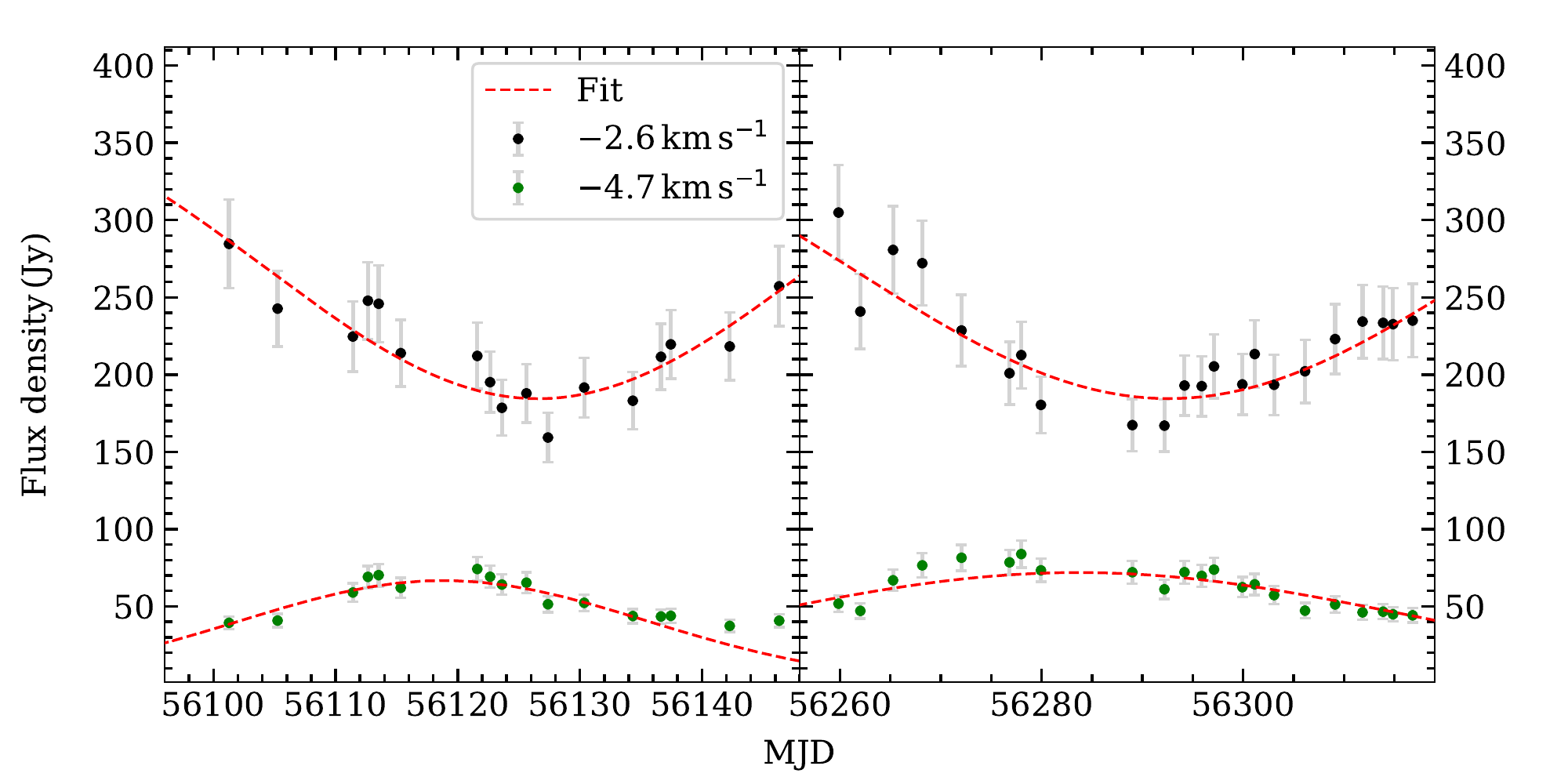}
\caption{Gaussian fits to light curves of the $-$2.6 and $-$4.7\kms\, features for periods MJD 56100-56140 and 56260-56300  \label{fig:anticor_gauss_fits} }
\end{figure}

Closer examination of the light curves in Figure\,\ref{fig:cepa_lcs} reveals anticorrelation between peak flux densities of the $-$2.6, $-$4.1, and $-$4.7\kms\, features during high-variability periods of the $-$4.7\kms\, component. To quantify this anticorrelation, we calculated Pearson correlation coefficients using the method PEARSONR from the SCIPY.STATS package \citep{2020SciPy-NMeth}. A similar analysis was performed by \citet{szymczak2014}; here we repeat it with the extended dataset. 
Results are presented in Figure\,\ref{fig:cepa-anticorr}. We find a statistically meaningful anticorrelation between light curves of the $-$2.6 and $-$4.7\kms\, features for both periods of activity of the $-$4.7\kms\, feature (MJD\,56057-56347 and MJD 57862-58851), but only for the first
period (MJD\,56057-56347) between $-$2.6 and $-$4.1\kms\,. There is also a statistically meaningful correlation between the features at $-$4.1 and $-$4.7\kms\,. The relation between peak flux densities of the $-$4.7 and $-$2.6\kms\, features appears to be non-linear in the second period (MJD 57862-58851), which suggests that the degree of saturation of one of the features varies during these periods, implying changes in the pumping efficiency.
We also calculated correlation coefficients for the whole observing period; the results are presented in Table \ref{tab:corr-params}.
\begin{table}
\centering
\caption{Correlation coefficients of the 6.7\,GHz spectral features in Cep\,A\,HW2 for the whole monitoring period. \label{tab:corr-params}}
\begin{tabular}[c]{l|rrrr}
\hline
Feature & \\
(\kms)  & $-$1.8 & $-$2.6 & $-$3.7 & $-$4.1 \\
\hline
\noalign{\vskip 0.5mm}  
$-$2.6  &    0.87 &          &              \\
$-$3.7 &    0.26 &    0.28 &              \\
$-$4.1  & 0.47 &    0.59 & 0.60         \\
$-$4.7 & $-$0.32 & $-$0.43 & 0.32 & 0.24 \\
\hline
\end{tabular}

\end{table}
The strong anticorrelation between the features at $-$2.6 and $-$4.7\kms\,  suggests that time-delays between variations of these features are relatively low. These time-delays likely correspond to the time it would take the light to travel between cloudlets. From \citet{sanna2017}, we can estimate the distance to 860\,AU, which would take about five days for the light to travel. We decided to fit the Gaussian function to visible symmetrical peaks in $-$4.7 and flux minima in $-$2.6\kms\, light curves to estimate time-lags. Estimated time-lags between the features at $-$4.7 and $-$2.6\kms\,  are 7.7$\pm$1.5 and 8.9$\pm$2.1 days for profiles \#1 and \#2, respectively. These values are well within an order of magnitude of the light-crossing time between $-$2.6 and $-$4.7\kms\, cloudlets (Table\,\ref{tab:gaussian_fits}, Fig.\,\ref{fig:anticor_gauss_fits}).
\begin{table}
\centering
\caption{Results of Gaussian fits shown in Fig.\,\ref{fig:anticor_gauss_fits}.\label{tab:gaussian_fits} }
\begin{tabular}[c]{c|ccc}
\hline
Profile & V$_{\mathrm{lsr}}$& MJD$_{\mathrm{peak}}$ & FWHM\\
No. & (\kms) & (d) & (d) \\
\hline
\multirow{2}{*}{1} & $-$2.6 & 56126.6 $\pm$ 1.1 & 52.4 $\pm$ 3.5 \\
                   & $-$4.7& 56118.8 $\pm$ 1.1 & 39.3 $\pm$ 2.3\\
\hline
\multirow{2}{*}{2} & $-$2.6 & 56292.6 $\pm$ 1.3 & 74.2 $\pm$ 4.7 \\
                   & $-$4.7 & 56283.7 $\pm$ 1.8 & 78.3 $\pm$ 4.7 \\
\hline
\end{tabular}

\end{table}

\begin{figure}
\centering
\includegraphics[width=0.9\columnwidth]{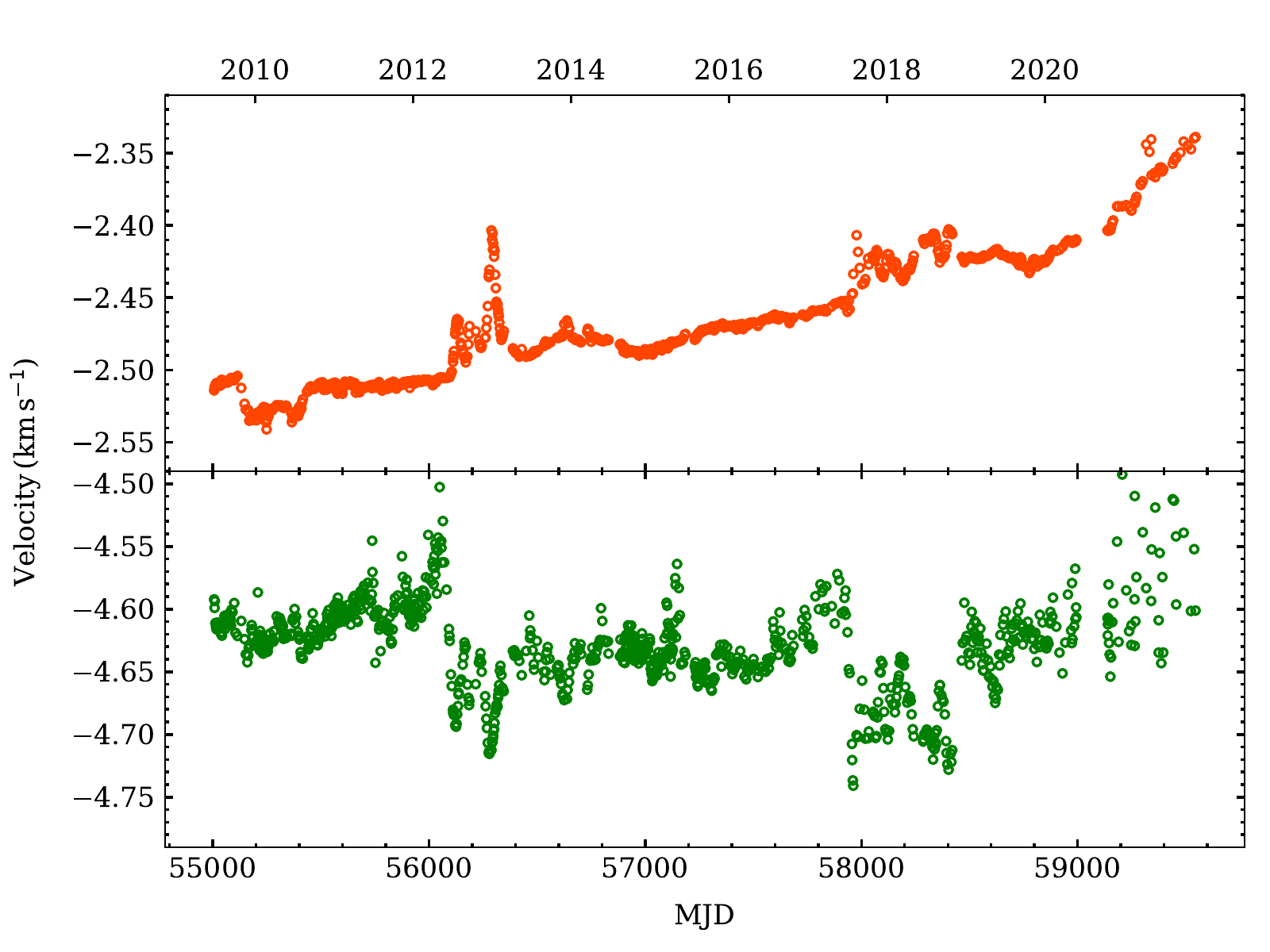}
\caption{Radial velocity curves for $-$2.6 and $-$4.7\kms\, features. \label{fig:radial_vel_curves_2p6-4p6}}
\end{figure}

Given that the features at $-$2.6 and $-$4.7\kms\,  show strong anticorrelation in intervals of MJD\,56057-56347 and MJD 57862-58851 (Fig.\,\ref{fig:cepa-anticorr}), it seems reasonable to examine how the peak velocities ($V_\mathrm{lsr}$) behave over time; results are presented in Figure\,\ref{fig:radial_vel_curves_2p6-4p6}.  In addition to an apparent non-linear drift towards positive $V_\mathrm{lsr}$, the feature at $-$2.6\kms shows rapid drifts towards positive $V_{\mathrm{lsr}}$ during the anticorrelation periods. The feature at $-$4.7\kms\,  shows the opposite behaviour; in specified periods, it drifts towards more negative values of $V_{\mathrm{lsr}}$.

\subsection{Properties of other regions}
Results of closer examination of cloudlets with Gaussian function fits are listed in Table\,\ref{tab:cloudlet_gauss_fit_results}. Their general characteristics are a simple linear or slightly arched morphology and an explicit velocity gradient. The size measured as the distance of the extreme spots ranges from 0.4 to 13\,AU, while the brightness temperature is between 
1.1$\times10^7$\, and 5.6$\times10^{10}$\,K.

\section{Discussion}
\subsection{Evidence of recurrent variability}
The first report of flares in  two redshifted features ($-$1.3 and $-$0.5\kms) at the beginning of 2010 was presented by \citet{szymczak2014}.
The present single-dish study indicates that the emission of these two features occurs regularly in intervals of 4.9 to 5.1 years. A comparison of our VLBI maps with those published by \citet{torstensson_2011}, \citet{sugiyama_2014}, and \citet{sanna2017} (figs\,\ref{fig:folded_light_curve} and \ref{fig:cluster_I_and_II_8_epochs}) provides a more detailed picture of the 6.7\,GHz maser variability in cluster II. We can see that the epoch 1 (Table\,\ref{tab:evn_projs}) observation  was made before the flare of the redshifted cloudlets, as our ephemerid suggests, and only the blueshifted spots ($-$2.5\kms) were detected. The observation in epoch 2 was conducted at the end of the flare, and faint (F$_{\mathrm{peak}}<$\,3.5\,Jy\,beam$^{-1}$) emission of the redshifted ($-$0.5, $-$1.3\kms) and blueshifted ($-$2.5\kms) cloudlets was seen. No emission at $-$0.5 and $-$1.3\kms\, was detected in epochs 3 or 4, that is, during a quiescent state anticipated by the ephemerid, and only faint blueshifted emission ($-$2.5\kms) was detected in epoch 3. In epochs 5 and 6, the EVN detected only faint blueshifted ($-$2.5\kms) cloudlets, while in epoch 7, they significantly increased, and the $-$0.5 and $-$1.3\kms\, cloudlets reappeared. In epoch 8, only the redshifted emission was seen for both cloudlets. In epoch 9, the emission at $-$0.5\kms\, diminished, whereas the $-$1.3\kms\, emission disappeared. We conclude that the published VLBI data fully confirm the recurrent appearance of the $-$1.3 and $-$0.5\kms\, emission, supporting our single-dish monitoring. Furthermore, these data may suggest that slightly blueshifted emission of $-$2.5\kms\, also exhibits a periodic behaviour. We note that this blueshifted emission in cluster II is highly blended with cluster I, preventing us from finding its light curve. 

\subsection{Flare characteristics and their origin}
Fast growth and a slow drop in the flux density are the main characteristics of the flare profile of the redshifted emission in Cep\,A\,HW2, and are preserved from cycle to cycle. The peak flux varies by a factor of 3,  while the duty cycle ranges from 0.06 to 0.16 (Fig.\,\ref{fig:folded_light_curve}). Here, the duty cycle is the fraction of a flare period for which the flux density is above half of the peak value. The rise-to-decay time ratio of the flare is 0.18 on average, and the relative amplitude is higher than 10. For a sample of all known periodic masers listed in \citet{olech2019} (their table A3), we infer that the duty cycle ranges from 0.16 to 0.97, while the  rise-to-decay time ratio is 0.3$-$2.0. The relative amplitude ranges from 0.2 to 5 in most cases, but there are five objects with lower limits of 5$-$120. Therefore, we find that the above-mentioned characteristics of the flared cloudlets in the target are close to the boundary values reported in known periodic masers. A period of 1828\,d is the only characteristic of Cep\,A\,HW2, which stands out by an order of magnitude from the median value of 200\,d \citep{olech2019} inferred for a sample of 25 known periodic masers.

There are three models that could explain the specific flare profile of the redshifted emission; pulsation instability due to the $\kappa$ mechanism \citep{inayoshi2013}, a colliding wind binary \citep{van_der_walt2011}, or spiral shocks in the central gap of the accretion disc \citep{parfenov2014}. Pulsation instability due to the $\kappa$ mechanism allows the stellar masses to be estimated with a power-law relation (\citealt{inayoshi_letter_2013}, their eq.\,2). 
In the case of Cep\,A\,HW2, a period of $\sim$\enspace5 years implies a mass of over 40\,M$_{\odot}$, which is between two and four times higher than values derived from the observed bolometric luminosity \citep{sanna2017}. We therefore argue that the observed variability cannot be explained by the $\kappa$ instability mechanism.

\citet{van_der_walt2011} explains the periodic variability of the methanol masers with a colliding wind binary scenario. Briefly,  periodic stellar wind interactions in the binary system generate additional free-free radiation that increases seed photon flux. Cooling of the plasma is much less abrupt, resulting in a characteristic profile with the rapid rise and slow decay. A rough check shows that both $-$0.5 and $-$1.3\kms\, maser cloudlets in cluster II lie very close to the radio continuum emission knots \citep{curiel2006} suggesting that the maser flares might be induced by modulation of the background radiation. The position of the centroid of the $-$0.5\kms\, maser emission measured with the EVN beam (Table\,\ref{tab:evn_projs}) in epochs 8 and 9 coincides within 45\,mas with the axis of radio jet \citep{curiel2006,carrasco2021}. A very similar coincidence is seen for the centroid of the $-$1.3\kms\, emission. The width of the 22\,GHz radio knot is $\sim$\enspace70\,AU \citep{curiel2006,carrasco2021}, while the size of the maser region in cluster II is 75\,AU (as estimated from epoch 7);  it is therefore quite certain that the maser volume could be illuminated by the background continuum radiation. Our observations imply that the brightness temperature difference between the active and quiet states of the $-$0.5 and $-$1.3\kms\, emission is more than one order of magnitude.  Cyclic changes of the continuum emission could have been induced by an accretion event or outflow. The radio continuum emission at centimetre wavelengths in Cep\,A\,HW2 shows moderate ($\sim$\enspace70\%) variability on a timescale of 10\,yr \citep{curiel2006}, which could be considered as a possible explanation for the characteristics of the flares. However, if the maser flares were generated by variations of background photon flux then there would be no time-lags (Table\,\ref{tab:flares}), which, on the other hand, we clearly observe. We therefore conclude that changes in the background radiation are not the cause of the maser flares in cluster II.

Characteristic flare profile could also be achieved by the spiral shock model \citep{parfenov2014}. In this scenario, an OB binary on a circular orbit creates shocks that travel through the central disc gap. Shocks interact with denser matter of the disc, effectively increasing IR radiation, which boosts the pumping of the methanol masers. This model is strongly constrained by the geometry of the system (which is required to be in an almost edge-on projection) and could also generate time-lags between the peaks of flaring cloudlets.  A chain of absorption and re-emission could slow down a pumping factor to subluminal velocity, but its magnitude is arguable. In section \ref{sec:pos-prop-red} we estimate the velocity of the pumping factor to $\sim$\enspace1000\kms, while the {heatwave} postulated by \citet{burns2020} penetrates a disc plane much faster, with a significant fraction of the speed of light. 

Comparison of the VLBI maser observations with continuum data \citep{patel2005} reveals that cluster II originates near the edge of a dust disc. The exposition of the masing cloudlets to the pumping IR photons from the disc could magnify the variability of the maser emission: an increase in the dust temperature boosts the flux of the pumping photons, which would increase the population-inverted abundance in cluster II, effectively raising the intensity of the masing cloudlets. Dust heats quickly \citep{johnstone2013}, which, combined with the relaxation time of the  maser, could result in the observed light curves. Previous studies \citep{cunningham2009} revealed that the Cep\,A\,HW2 system hosts many companion stars; thus, it is probable that an increase in dust temperature could be related to a companion star, modulating the accretion rate.

Using the NEOWISE survey \citep{neowise_paper}, we might check whether or not the IR flux changes over time and whether or not it correlates with the 6.7\,GHz maser emission; \citet{olech2020, olech2019} did this for periodic sources G107.298+5.639 and G59.533$-$0.192. However, Cep\,A\,HW2 is a crowded region, and WISE blends several sources. We therefore conclude that a higher angular resolution IR monitoring is required to resolve this case.
This scenario also does not fully explain the observed time-lags. The resulting chain of absorption and re-emission (heatwave) should propagate with a velocity of several hundred \kms\, (see section \ref{sec:pos-prop-red}), which is too slow compared to the reported case \citep{burns2020}. We note that the 2010 flare has been successfully modelled as an effect of Dicke's superradiance \citep{rajabi2019}: in this scenario, flaring behaviour triggers when the critical inverted population column density is exceeded. This effect might impact the observed delays.

One might expect a driving pulse ---induced for instance by the accretion luminosity burst--- to affect all the maser cloudlets in Cep\,A\,HW2, but the response of different parts of the molecular disc is very different. One possible cause could be differences in the degree of saturation in maser cloudlets. Following Eq.\,1 of \citealt{vlemmings2010},   we attempt to estimate the ratio of the maser-stimulated emission rate $R$ to the maser decay rate $\Gamma,$ which is the measure of saturation level. For the beaming solid angle $\Delta \Omega \simeq 10^{-2}$, $\Gamma$ = 1\,s$^{-1}$ and the maser brightness temperature from Tables\,\ref{tab:spots_rd002} and \ref{tab:spots_ed048B}, we obtain a saturation level ranging from 0.005 to 0.2 for most of the cloudlets composing the persistent features.  For the $-$0.5 and $-$4.7\kms\, cloudlets, $R/\Gamma$ is $8\times10^{-4}$ and $2\times10^{-4}$, respectively. 
It is therefore likely that the varying behaviour of the maser features, in response to the same driving factor, is related to the degree of saturation. 

\subsection{Causes of velocity drifts}
\begin{figure}
\centering
\includegraphics[width=1\columnwidth]{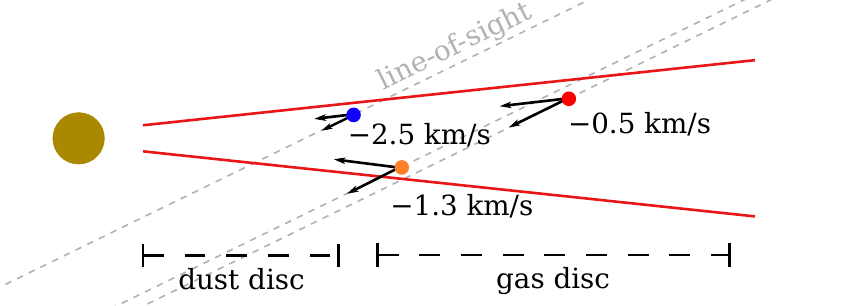}
\caption{Cartoon to illustrate the proposed scenario for infall motion in cluster II. The coloured circles denote the flaring cloudlets and the labels correspond to their velocities along the line of sight. \label{fig:cepa_infall_scheme}}
\end{figure}

The drift in the radial velocity of the redshifted emission is a particularly interesting phenomenon, showing a linear trend with the rate of 3-4\,$\times$\,10$^{-5}$\kms\,d$^{-1}$ for about 11.5\,yr of our monitoring. It is striking that this rate is almost the same as that reported in the archetypal periodic source G9.62+0.20E \citep{macleod2021}. The profile of each of the two redshifted features is satisfactorily fitted with a single Gaussian for both the 32\,m dish and EVN spectra; therefore we exclude a possibility that the observed systematic drifts are artifacts produced by variability of two or more spectral components with very close velocities. \citet{macleod2021} proposed two hypotheses to explain the velocity drifts; (i) precession of a Keplerian disc and (ii) infall motion. Adopting the disc model of Cep\,A\,HW2 from \citet{sanna2017} and using the maser distribution from EVN observations (Fig.\,\ref{fig:cepa-spots-june-and-october-2020}), in line with the approach of MacLeod et al., we obtain a precession period of 870\,yr. This value is more than an order of magnitude shorter than a putative period ($\sim$\enspace8\,$\times$\,10$^4$\,yr) of disc precession suggested from H$_2$ outflow observations \citep{cunningham2009}, and therefore hypothesis (i) can be ruled out. 
There is observational evidence for infall motion of 6.7\,GHz maser cloudlets with a velocity of $\sim$\enspace2\kms\, in a plane of the accretion disc \citep{torstensson_2011, sanna2017}. Our observations confirm the findings of these latter authors and suggest infall of the most redshifted cloudlets with a velocity of 4\kms\, (Sect.\ref{sec:pos-prop-red}). 
As circumstellar discs reach number densities higher than 10$^9$cm$^{-3}$ \citep{vlemmings2010}, the maser emission is unlikely to originate from the densest inner part of the disc. Thus, we propose a scenario (Fig.\,\ref{fig:cepa_infall_scheme}) in which the maser clouds are located near the upper and lower edges of the disc and follow infall motions directed into HW2, parallel to the edges. With this scheme, the radial velocity drift (Fig.\,\ref{fig:radial_curve}) is due to cloudlets accelerating towards HW2. Furthermore, this picture is entirely consistent with the measured time-lags between the onset of the flares of $-$0.5 and $-$1.3\kms\, cloudlets (Table\,\ref{tab:flares}).

\subsection{Anticorrelated variability}
The synchronous anticorrelation between the features at $-$1.8, $-$2.6, $-$3.7, $-$4.1, and $-$4.7\kms\,  was reported for the first time by \citet{sugiyama_2008}. They monitored the target over $\sim$\enspace3\,months and found that since MJD 54320, the flux density of the feature at $-$4.9\kms\,  had increased by a factor of 3 over a timescale of 2.5\,weeks, while that of $-$2.7\kms\, decreased by a factor of 1.5. \citet{szymczak2014} revealed that the $-$4.7\kms\, feature increased in intensity by a factor of 8 since MJD 56118. Our new data show another rapid five-fold increase of its amplitude since MJD 57960 (Fig.\,\ref{fig:cepa_lcs}). It is striking that the event of anticorrelated and synchronous variability reported by \citet{szymczak2014} occurred 1798\,d after that noticed by \citet{sugiyama_2008} and 1842\,d before that observed here. These time intervals are very well in line with a periodicity of flares of the redshifted emission.

Figure \ref{fig:cepa-anticorr} displays the flux densities of the features at $-$2.6, $-$4.1, and $-$4.7\kms\,  plotted against each other. Red points originate from the period of MJD\,56057$-$56347, which is almost the same as in \citet{szymczak2014} (their fig. 4), and the green points come from our new observations of MJD 57862$-$58851. There is a linear anticorrelation between the features at $-$2.6 and $-$4.7\kms\,  in the first time-span, while we note a significant deviation from a linear relation in a second one. This suggests that the causes of this behaviour affect the optical depth of the  masing cloud, resulting in a non-linear relation. There are also visible radial velocity drifts during these periods of activity (Fig.\,\ref{fig:radial_vel_curves_2p6-4p6}). A recent study \citep{sanna2017} indicated that the velocity field in Cep\,A\,HW2 has a Keplerian component; it therefore becomes possible to explain the anticorrelated variability with the scenario proposed by \citet{elmegreen1979,cesaroni1990,szymczak2014}: when the pump rate drops on one side of the Keplerian disc, the opposite side of it will be affected soon after the light-crossing time, resulting in synchronous anticorrelated changes in the flux density of red- and blueshifted emission. This scenario requires a nearly edge-on oriented disc, but the inclination angle is $\sim$\enspace64\degr\, \citep{sanna2017}  in Cep\,A\,HW2, and this condition is only partially fulfilled.

\section{Conclusions}
We report the detection of the quasi-periodic high-relative-amplitude flares of the extremely redshifted (V$_{\mathrm{lsr}}\,>\,-$1.5\kms) 6.7\,GHz methanol maser emission in Cep\,A\,HW2. The flares occur at $\sim$\enspace1800-day intervals and originate from the maser cluster closest to the presumed position of the protostar, near the edge of the dust disc. The spatial distribution of the flaring cloudlets, the radial velocity drift, and the time-lags of the flare peaks are consistent with a scenario of a molecular disc rotating with subKeplerian velocity and an infall motion component. Presumably, periodic and minor changes in the luminosity of the powering system result in significant variations of emission from unsaturated maser regions at the extreme redshifted velocities. Still, these changes do not  significantly affect the most persistent maser emission regions approaching saturation state. The measured time-lags imply that changes in physical conditions that cause maser flares propagate at $\sim$\enspace1000\kms. 
Evidence of time-lags is against the alternative hypothesis that the flaring activity is caused by the amplification of variable background radiation from the radio jet. Synchronous anticorrelated changes in the flux densities of the $-$2.6 and $-$4.7\kms\, spectral components occur at approximately five-year intervals and suggest radiative interaction between the different parts of the molecular disc. Further long-term studies of the target are needed to confirm the maser periodicity and explain its origin.

\section{Acknowledgements}
The 32\,m radio telescope is operated by the Institute Astronomy, Nicolaus Copernicus University and supported by the Polish Ministry of Science and Higher Education SpUB grant. We thank the staff and students for assistance with the observations.
The research has made use of the SIMBAD data base, operated at CDS (Strasbourg, France), as well as NASA's Astrophysics Data System Bibliographic Services. This research made use of Astropy\footnote{http://www.astropy.org}, a community-developed core Python package for Astronomy \citep{astropy:2013, astropy:2018}. M.O. thanks the Ministry of Education and Science of the Republic of Poland for support and granting funds for the Polish contribution to the International LOFAR Telescope (arrangement no 2021/WK/02) and for maintenance of the LOFAR PL-612 Baldy (MSHE decision no. 59/E-383/SPUB/SP/2019.1)

\bibliography{librarian}{}

\begin{thebibliography}{60}
\expandafter\ifx\csname natexlab\endcsname\relax\def\natexlab#1{#1}\fi

\bibitem[{{Astropy Collaboration} {et~al.}(2018){Astropy Collaboration},
  {Price-Whelan}, {Sip{\H{o}}cz}, {G{\"u}nther}, {Lim}, {Crawford}, {Conseil},
  {Shupe}, {Craig}, {Dencheva}, {Ginsburg}, {Vand erPlas}, {Bradley},
  {P{\'e}rez-Su{\'a}rez}, {de Val-Borro}, {Aldcroft}, {Cruz}, {Robitaille},
  {Tollerud}, {Ardelean}, {Babej}, {Bach}, {Bachetti}, {Bakanov}, {Bamford},
  {Barentsen}, {Barmby}, {Baumbach}, {Berry}, {Biscani}, {Boquien}, {Bostroem},
  {Bouma}, {Brammer}, {Bray}, {Breytenbach}, {Buddelmeijer}, {Burke},
  {Calderone}, {Cano Rodr{\'\i}guez}, {Cara}, {Cardoso}, {Cheedella}, {Copin},
  {Corrales}, {Crichton}, {D'Avella}, {Deil}, {Depagne}, {Dietrich}, {Donath},
  {Droettboom}, {Earl}, {Erben}, {Fabbro}, {Ferreira}, {Finethy}, {Fox},
  {Garrison}, {Gibbons}, {Goldstein}, {Gommers}, {Greco}, {Greenfield},
  {Groener}, {Grollier}, {Hagen}, {Hirst}, {Homeier}, {Horton}, {Hosseinzadeh},
  {Hu}, {Hunkeler}, {Ivezi{\'c}}, {Jain}, {Jenness}, {Kanarek}, {Kendrew},
  {Kern}, {Kerzendorf}, {Khvalko}, {King}, {Kirkby}, {Kulkarni}, {Kumar},
  {Lee}, {Lenz}, {Littlefair}, {Ma}, {Macleod}, {Mastropietro}, {McCully},
  {Montagnac}, {Morris}, {Mueller}, {Mumford}, {Muna}, {Murphy}, {Nelson},
  {Nguyen}, {Ninan}, {N{\"o}the}, {Ogaz}, {Oh}, {Parejko}, {Parley}, {Pascual},
  {Patil}, {Patil}, {Plunkett}, {Prochaska}, {Rastogi}, {Reddy Janga},
  {Sabater}, {Sakurikar}, {Seifert}, {Sherbert}, {Sherwood-Taylor}, {Shih},
  {Sick}, {Silbiger}, {Singanamalla}, {Singer}, {Sladen}, {Sooley},
  {Sornarajah}, {Streicher}, {Teuben}, {Thomas}, {Tremblay}, {Turner},
  {Terr{\'o}n}, {van Kerkwijk}, {de la Vega}, {Watkins}, {Weaver}, {Whitmore},
  {Woillez}, {Zabalza}, \& {Astropy Contributors}}]{astropy:2018}
{Astropy Collaboration}, {Price-Whelan}, A.~M., {Sip{\H{o}}cz}, B.~M., {et~al.}
  2018, \aj, 156, 123

\bibitem[{{Astropy Collaboration} {et~al.}(2013){Astropy Collaboration},
  {Robitaille}, {Tollerud}, {Greenfield}, {Droettboom}, {Bray}, {Aldcroft},
  {Davis}, {Ginsburg}, {Price-Whelan}, {Kerzendorf}, {Conley}, {Crighton},
  {Barbary}, {Muna}, {Ferguson}, {Grollier}, {Parikh}, {Nair}, {Unther},
  {Deil}, {Woillez}, {Conseil}, {Kramer}, {Turner}, {Singer}, {Fox}, {Weaver},
  {Zabalza}, {Edwards}, {Azalee Bostroem}, {Burke}, {Casey}, {Crawford},
  {Dencheva}, {Ely}, {Jenness}, {Labrie}, {Lim}, {Pierfederici}, {Pontzen},
  {Ptak}, {Refsdal}, {Servillat}, \& {Streicher}}]{astropy:2013}
{Astropy Collaboration}, {Robitaille}, T.~P., {Tollerud}, E.~J., {et~al.} 2013,
  \aap, 558, A33

\bibitem[{{Breen} {et~al.}(2015){Breen}, {Fuller}, {Caswell}, {Green},
  {Avison}, {Ellingsen}, {Gray}, {Pestalozzi}, {Quinn}, {Richards}, {Thompson},
  \& {Voronkov}}]{breen2015}
{Breen}, S.~L., {Fuller}, G.~A., {Caswell}, J.~L., {et~al.} 2015, \mnras, 450,
  4109

\bibitem[{{Burns} {et~al.}(2020){Burns}, {Sugiyama}, {Hirota}, {Kim},
  {Sobolev}, {Stecklum}, {MacLeod}, {Yonekura}, {Olech}, {Orosz}, {Ellingsen},
  {Hyland}, {Caratti o Garatti}, {Brogan}, {Hunter}, {Phillips}, {van den
  Heever}, {Eisl{\"o}ffel}, {Linz}, {Surcis}, {Chibueze}, {Baan}, \&
  {Kramer}}]{burns2020}
{Burns}, R.~A., {Sugiyama}, K., {Hirota}, T., {et~al.} 2020, Nature Astronomy,
  4, 506

\bibitem[{{Carrasco-Gonz{\'a}lez} {et~al.}(2021){Carrasco-Gonz{\'a}lez},
  {Sanna}, {Rodr{\'\i}guez-Kamenetzky}, {Moscadelli}, {Hoare}, {Torrelles},
  {Galv{\'a}n-Madrid}, \& {Izquierdo}}]{carrasco2021}
{Carrasco-Gonz{\'a}lez}, C., {Sanna}, A., {Rodr{\'\i}guez-Kamenetzky}, A.,
  {et~al.} 2021, \apjl, 914, L1

\bibitem[{{Cesaroni}(1990)}]{cesaroni1990}
{Cesaroni}, R. 1990, \aap, 233, 513

\bibitem[{{Cragg} {et~al.}(2005){Cragg}, {Sobolev}, \& {Godfrey}}]{cragg2005}
{Cragg}, D.~M., {Sobolev}, A.~M., \& {Godfrey}, P.~D. 2005, \mnras, 360, 533

\bibitem[{{Cunningham} {et~al.}(2009){Cunningham}, {Moeckel}, \&
  {Bally}}]{cunningham2009}
{Cunningham}, N.~J., {Moeckel}, N., \& {Bally}, J. 2009, \apj, 692, 943

\bibitem[{{Curiel} {et~al.}(2006){Curiel}, {Ho}, {Patel}, {Torrelles},
  {Rodr{\'\i}guez}, {Trinidad}, {Cant{\'o}}, {Hern{\'a}ndez}, {G{\'o}mez},
  {Garay}, \& {Anglada}}]{curiel2006}
{Curiel}, S., {Ho}, P.~T.~P., {Patel}, N.~A., {et~al.} 2006, \apj, 638, 878

\bibitem[{{Durjasz} {et~al.}(2019){Durjasz}, {Szymczak}, \&
  {Olech}}]{durjasz2019}
{Durjasz}, M., {Szymczak}, M., \& {Olech}, M. 2019, \mnras, 485, 777

\bibitem[{{Elmegreen} \& {Morris}(1979)}]{elmegreen1979}
{Elmegreen}, B.~G. \& {Morris}, M. 1979, \apj, 229, 593

\bibitem[{{Fontani} {et~al.}(2010){Fontani}, {Cesaroni}, \&
  {Furuya}}]{fontani_2010}
{Fontani}, F., {Cesaroni}, R., \& {Furuya}, R.~S. 2010, \aap, 517, A56

\bibitem[{{Fujisawa} {et~al.}(2014){Fujisawa}, {Takase}, {Kimura}, {Aoki},
  {Nagadomi}, {Shimomura}, {Sugiyama}, {Motogi}, {Niinuma}, {Hirota}, \&
  {Yonekura}}]{fujisawa2014}
{Fujisawa}, K., {Takase}, G., {Kimura}, S., {et~al.} 2014, \pasj, 66, 78

\bibitem[{{Gallimore} {et~al.}(2003){Gallimore}, {Cool}, {Thornley}, \&
  {McMullin}}]{gallimore_2003}
{Gallimore}, J.~F., {Cool}, R.~J., {Thornley}, M.~D., \& {McMullin}, J. 2003,
  \apj, 586, 306

\bibitem[{{Goedhart} {et~al.}(2003){Goedhart}, {Gaylard}, \& {van der
  Walt}}]{goedhart2003}
{Goedhart}, S., {Gaylard}, M.~J., \& {van der Walt}, D.~J. 2003, \mnras, 339,
  L33

\bibitem[{{Goedhart} {et~al.}(2004){Goedhart}, {Gaylard}, \& {van der
  Walt}}]{goedhart2004}
{Goedhart}, S., {Gaylard}, M.~J., \& {van der Walt}, D.~J. 2004, \mnras, 355,
  553

\bibitem[{{Goedhart} {et~al.}(2009){Goedhart}, {Langa}, {Gaylard}, \& {Van Der
  Walt}}]{goedhart2009}
{Goedhart}, S., {Langa}, M.~C., {Gaylard}, M.~J., \& {Van Der Walt}, D.~J.
  2009, \mnras, 398, 995

\bibitem[{{Green} {et~al.}(2010){Green}, {Caswell}, {Fuller}, {Avison},
  {Breen}, {Ellingsen}, {Gray}, {Pestalozzi}, {Quinn}, {Thompson}, \&
  {Voronkov}}]{green2010}
{Green}, J.~A., {Caswell}, J.~L., {Fuller}, G.~A., {et~al.} 2010, \mnras, 409,
  913

\bibitem[{{Hu} {et~al.}(2016){Hu}, {Menten}, {Wu}, {Bartkiewicz}, {Rygl},
  {Reid}, {Urquhart}, \& {Zheng}}]{hu2016}
{Hu}, B., {Menten}, K.~M., {Wu}, Y., {et~al.} 2016, \apj, 833, 18

\bibitem[{{Hughes} \& {Wouterloot}(1984)}]{hw_paper}
{Hughes}, V.~A. \& {Wouterloot}, J.~G.~A. 1984, \apj, 276, 204

\bibitem[{{Hunter} {et~al.}(2018){Hunter}, {Brogan}, {MacLeod}, {Cyganowski},
  {Chibueze}, {Friesen}, {Hirota}, {Smits}, {Chandler}, \&
  {Indebetouw}}]{hunter2018}
{Hunter}, T.~R., {Brogan}, C.~L., {MacLeod}, G.~C., {et~al.} 2018, \apj, 854,
  170

\bibitem[{{Inayoshi} {et~al.}(2013{\natexlab{a}}){Inayoshi}, {Hosokawa}, \&
  {Omukai}}]{inayoshi2013}
{Inayoshi}, K., {Hosokawa}, T., \& {Omukai}, K. 2013{\natexlab{a}}, \mnras,
  431, 3036

\bibitem[{{Inayoshi} {et~al.}(2013{\natexlab{b}}){Inayoshi}, {Sugiyama},
  {Hosokawa}, {Motogi}, \& {Tanaka}}]{inayoshi_letter_2013}
{Inayoshi}, K., {Sugiyama}, K., {Hosokawa}, T., {Motogi}, K., \& {Tanaka}, K.
  E.~I. 2013{\natexlab{b}}, \apjl, 769, L20

\bibitem[{{Johnstone} {et~al.}(2013){Johnstone}, {Hendricks}, {Herczeg}, \&
  {Bruderer}}]{johnstone2013}
{Johnstone}, D., {Hendricks}, B., {Herczeg}, G.~J., \& {Bruderer}, S. 2013,
  \apj, 765, 133

\bibitem[{{Keimpema} {et~al.}(2015){Keimpema}, {Kettenis}, {Pogrebenko},
  {Campbell}, {Cim{\'o}}, {Duev}, {Eldering}, {Kruithof}, {van Langevelde},
  {Marchal}, {Molera Calv{\'e}s}, {Ozdemir}, {Paragi}, {Pidopryhora},
  {Szomoru}, \& {Yang}}]{keimpema2015}
{Keimpema}, A., {Kettenis}, M.~M., {Pogrebenko}, S.~V., {et~al.} 2015,
  Experimental Astronomy, 39, 259

\bibitem[{{Ladeyschikov} {et~al.}(2019){Ladeyschikov}, {Bayandina}, \&
  {Sobolev}}]{maserdb_paper}
{Ladeyschikov}, D.~A., {Bayandina}, O.~S., \& {Sobolev}, A.~M. 2019, \aj, 158,
  233

\bibitem[{{Lew}(2018)}]{lew2018}
{Lew}, B. 2018, Experimental Astronomy, 45, 81

\bibitem[{{MacLeod} {et~al.}(2021){MacLeod}, {Chibueze}, {Sanna}, {Paulsen},
  {Houde}, {van den Heever}, \& {Goedhart}}]{macleod2021}
{MacLeod}, G.~C., {Chibueze}, J.~O., {Sanna}, A., {et~al.} 2021, \mnras, 500,
  3425

\bibitem[{{Mainzer} {et~al.}(2011){Mainzer}, {Bauer}, {Grav}, {Masiero},
  {Cutri}, {Dailey}, {Eisenhardt}, {McMillan}, {Wright}, {Walker}, {Jedicke},
  {Spahr}, {Tholen}, {Alles}, {Beck}, {Brandenburg}, {Conrow}, {Evans},
  {Fowler}, {Jarrett}, {Marsh}, {Masci}, {McCallon}, {Wheelock}, {Wittman},
  {Wyatt}, {DeBaun}, {Elliott}, {Elsbury}, {Gautier}, {Gomillion}, {Leisawitz},
  {Maleszewski}, {Micheli}, \& {Wilkins}}]{neowise_paper}
{Mainzer}, A., {Bauer}, J., {Grav}, T., {et~al.} 2011, \apj, 731, 53

\bibitem[{{Maswanganye} {et~al.}(2016){Maswanganye}, {van der Walt},
  {Goedhart}, \& {Gaylard}}]{maswanganye2016}
{Maswanganye}, J.~P., {van der Walt}, D.~J., {Goedhart}, S., \& {Gaylard},
  M.~J. 2016, \mnras, 456, 4335

\bibitem[{{Menten}(1991)}]{menten1991}
{Menten}, K.~M. 1991, \apjl, 380, L75

\bibitem[{{Moscadelli} {et~al.}(2009){Moscadelli}, {Reid}, {Menten},
  {Brunthaler}, {Zheng}, \& {Xu}}]{moscadelli2009}
{Moscadelli}, L., {Reid}, M.~J., {Menten}, K.~M., {et~al.} 2009, \apj, 693, 406

\bibitem[{{Moscadelli} {et~al.}(2017){Moscadelli}, {Sanna}, {Goddi},
  {Walmsley}, {Cesaroni}, {Caratti o Garatti}, {Stecklum}, {Menten}, \&
  {Kraus}}]{moscadelli2017}
{Moscadelli}, L., {Sanna}, A., {Goddi}, C., {et~al.} 2017, \aap, 600, L8

\bibitem[{{Olech} {et~al.}(2020){Olech}, {Szymczak}, {Wolak}, {G{\'e}rard}, \&
  {Bartkiewicz}}]{olech2020}
{Olech}, M., {Szymczak}, M., {Wolak}, P., {G{\'e}rard}, E., \& {Bartkiewicz},
  A. 2020, \aap, 634, A41

\bibitem[{{Olech} {et~al.}(2019){Olech}, {Szymczak}, {Wolak}, {Sarniak}, \&
  {Bartkiewicz}}]{olech2019}
{Olech}, M., {Szymczak}, M., {Wolak}, P., {Sarniak}, R., \& {Bartkiewicz}, A.
  2019, \mnras, 486, 1236

\bibitem[{{Pandian} {et~al.}(2007){Pandian}, {Goldsmith}, \& {Deshpand
  e}}]{pandian2007}
{Pandian}, J.~D., {Goldsmith}, P.~F., \& {Deshpand e}, A.~A. 2007, \apj, 656,
  255

\bibitem[{{Parfenov} \& {Sobolev}(2014)}]{parfenov2014}
{Parfenov}, S.~Y. \& {Sobolev}, A.~M. 2014, \mnras, 444, 620

\bibitem[{{Patel} {et~al.}(2005){Patel}, {Curiel}, {Sridharan}, {Zhang},
  {Hunter}, {Ho}, {Torrelles}, {Moran}, {G{\'o}mez}, \& {Anglada}}]{patel2005}
{Patel}, N.~A., {Curiel}, S., {Sridharan}, T.~K., {et~al.} 2005, \nat, 437, 109

\bibitem[{{Pietka} {et~al.}(2015){Pietka}, {Fender}, \& {Keane}}]{pietka2015}
{Pietka}, M., {Fender}, R.~P., \& {Keane}, E.~F. 2015, \mnras, 446, 3687

\bibitem[{{Rajabi} {et~al.}(2019){Rajabi}, {Houde}, {Bartkiewicz}, {Olech},
  {Szymczak}, \& {Wolak}}]{rajabi2019}
{Rajabi}, F., {Houde}, M., {Bartkiewicz}, A., {et~al.} 2019, \mnras, 484, 1590

\bibitem[{{Rodriguez} {et~al.}(1994){Rodriguez}, {Garay}, {Curiel}, {Ramirez},
  {Torrelles}, {Gomez}, \& {Velazquez}}]{rodrigues1994}
{Rodriguez}, L.~F., {Garay}, G., {Curiel}, S., {et~al.} 1994, \apjl, 430, L65

\bibitem[{{Sanna} {et~al.}(2017){Sanna}, {Moscadelli}, {Surcis}, {van
  Langevelde}, {Torstensson}, \& {Sobolev}}]{sanna2017}
{Sanna}, A., {Moscadelli}, L., {Surcis}, G., {et~al.} 2017, \aap, 603, A94

\bibitem[{{Sobolev} {et~al.}(1997){Sobolev}, {Cragg}, \&
  {Godfrey}}]{sobolev1997}
{Sobolev}, A.~M., {Cragg}, D.~M., \& {Godfrey}, P.~D. 1997, \aap, 324, 211

\bibitem[{{Sugiyama} {et~al.}(2008){Sugiyama}, {Fujisawa}, {Doi}, {Honma},
  {Isono}, {Kobayashi}, {Mochizuki}, \& {Murata}}]{sugiyama_2008}
{Sugiyama}, K., {Fujisawa}, K., {Doi}, A., {et~al.} 2008, \pasj, 60, 1001

\bibitem[{{Sugiyama} {et~al.}(2014){Sugiyama}, {Fujisawa}, {Doi}, {Honma},
  {Kobayashi}, {Murata}, {Motogi}, {Niinuma}, {Ogawa}, {Wajima},
  {Sawada-Satoh}, \& {Ellingsen}}]{sugiyama_2014}
{Sugiyama}, K., {Fujisawa}, K., {Doi}, A., {et~al.} 2014, \aap, 562, A82

\bibitem[{{Szymczak} {et~al.}(2000){Szymczak}, {Hrynek}, \&
  {Kus}}]{szymczak2000}
{Szymczak}, M., {Hrynek}, G., \& {Kus}, A.~J. 2000, \aaps, 143, 269

\bibitem[{{Szymczak} {et~al.}(2018){Szymczak}, {Olech}, {Sarniak}, {Wolak}, \&
  {Bartkiewicz}}]{szymczak2018}
{Szymczak}, M., {Olech}, M., {Sarniak}, R., {Wolak}, P., \& {Bartkiewicz}, A.
  2018, \mnras, 474, 219

\bibitem[{{Szymczak} {et~al.}(2016){Szymczak}, {Olech}, {Wolak}, {Bartkiewicz},
  \& {Gawro{\'n}ski}}]{szymczak_2016}
{Szymczak}, M., {Olech}, M., {Wolak}, P., {Bartkiewicz}, A., \&
  {Gawro{\'n}ski}, M. 2016, \mnras, 459, L56

\bibitem[{{Szymczak} {et~al.}(2014){Szymczak}, {Wolak}, \&
  {Bartkiewicz}}]{szymczak2014}
{Szymczak}, M., {Wolak}, P., \& {Bartkiewicz}, A. 2014, \mnras, 439, 407

\bibitem[{{Szymczak} {et~al.}(2012){Szymczak}, {Wolak}, {Bartkiewicz}, \&
  {Borkowski}}]{szymczak2012}
{Szymczak}, M., {Wolak}, P., {Bartkiewicz}, A., \& {Borkowski}, K.~M. 2012,
  Astronomische Nachrichten, 333, 634

\bibitem[{{Szymczak} {et~al.}(2011){Szymczak}, {Wolak}, {Bartkiewicz}, \& {van
  Langevelde}}]{szymczak2011}
{Szymczak}, M., {Wolak}, P., {Bartkiewicz}, A., \& {van Langevelde}, H.~J.
  2011, \aap, 531, L3

\bibitem[{{Torrelles} {et~al.}(1996){Torrelles}, {Gomez}, {Rodriguez},
  {Curiel}, {Ho}, \& {Garay}}]{torrelles_1996}
{Torrelles}, J.~M., {Gomez}, J.~F., {Rodriguez}, L.~F., {et~al.} 1996, \apjl,
  457, L107

\bibitem[{{Torrelles} {et~al.}(2011){Torrelles}, {Patel}, {Curiel},
  {Estalella}, {G{\'o}mez}, {Rodr{\'\i}guez}, {Cant{\'o}}, {Anglada},
  {Vlemmings}, {Garay}, {Raga}, \& {Ho}}]{torrelles_2011}
{Torrelles}, J.~M., {Patel}, N.~A., {Curiel}, S., {et~al.} 2011, \mnras, 410,
  627

\bibitem[{{Torrelles} {et~al.}(2001){Torrelles}, {Patel}, {G{\'o}mez}, {Ho},
  {Rodr{\'\i}guez}, {Anglada}, {Garay}, {Greenhill}, {Curiel}, \&
  {Cant{\'o}}}]{torrelles_2001}
{Torrelles}, J.~M., {Patel}, N.~A., {G{\'o}mez}, J.~F., {et~al.} 2001, \nat,
  411, 277

\bibitem[{{Torstensson} {et~al.}(2011){Torstensson}, {van Langevelde},
  {Vlemmings}, \& {Bourke}}]{torstensson_2011}
{Torstensson}, K.~J.~E., {van Langevelde}, H.~J., {Vlemmings}, W.~H.~T., \&
  {Bourke}, S. 2011, \aap, 526, A38

\bibitem[{{van der Walt}(2011)}]{van_der_walt2011}
{van der Walt}, D.~J. 2011, \aj, 141, 152

\bibitem[{Virtanen {et~al.}(2020)Virtanen, Gommers, Oliphant, Haberland, Reddy,
  Cournapeau, Burovski, Peterson, Weckesser, Bright, {van der Walt}, Brett,
  Wilson, Millman, Mayorov, Nelson, Jones, Kern, Larson, Carey, Polat, Feng,
  Moore, {VanderPlas}, Laxalde, Perktold, Cimrman, Henriksen, Quintero, Harris,
  Archibald, Ribeiro, Pedregosa, {van Mulbregt}, \& {SciPy 1.0
  Contributors}}]{2020SciPy-NMeth}
Virtanen, P., Gommers, R., Oliphant, T.~E., {et~al.} 2020, Nature Methods, 17,
  261

\bibitem[{{Vlemmings}(2008)}]{vlemmings_2008}
{Vlemmings}, W.~H.~T. 2008, \aap, 484, 773

\bibitem[{{Vlemmings} {et~al.}(2010){Vlemmings}, {Surcis}, {Torstensson}, \&
  {van Langevelde}}]{vlemmings2010}
{Vlemmings}, W.~H.~T., {Surcis}, G., {Torstensson}, K.~J.~E., \& {van
  Langevelde}, H.~J. 2010, \mnras, 404, 134

\bibitem[{{Yang} {et~al.}(2017){Yang}, {Chen}, {Shen}, {Li}, {Wang}, {Jiang},
  {Li}, {Dong}, {Wu}, {Qiao}, \& {Ren}}]{yang_2017}
{Yang}, K., {Chen}, X., {Shen}, Z.-Q., {et~al.} 2017, \apj, 846, 160

\end{thebibliography}
\bibliographystyle{aa}

\begin{appendix}

\onecolumn
\section{Spot tables}

\begin{longtable}{ p{0.15\textwidth}p{0.15\textwidth}p{0.15\textwidth}p{0.15\textwidth}p{0.15\textwidth}p{0.15\textwidth} }
\caption{List of 6.7 GHz CH$_3$OH maser spots detected in epoch 8.
\label{tab:spots_rd002} }
\\\hline
\noalign{\vskip 1mm}  
Cluster & $\Delta$\,$\alpha$ & $\Delta$\,$\delta$ & V$_{\mathrm{lsr}}$ & F$_{\mathrm{peak}}$  & T$_{\mathrm{b}}$ \\
 & (mas) & (mas) & (\kms) & (Jy\,beam$^{-1}$) & (K)
\\\hline
\noalign{\vskip 1mm}  
\endfirsthead
\caption{continued}
\\\hline
\noalign{\vskip 1mm}  
Cluster & $\Delta$\,$\alpha$ & $\Delta$\,$\delta$ & V$_{\mathrm{lsr}}$ & F$_{\mathrm{peak}}$  & T$_{\mathrm{b}}$ \\
 & (mas) & (mas) & (\kms) & (Jy\,beam$^{-1}$) & (K)
\\\hline
\noalign{\vskip 1mm}  
\endhead
\hline
\endfoot
\hline
\noalign{\vskip 1mm}
\multicolumn{6}{r}{\tablefoot{Numbers in the first column refer to the cluster that the spot originates from (see figure \ref{fig:cepa-spots-june-and-october-2020}). Columns 2 and 3 give the relative positions and uncertainties in Right Ascension and Declination, respectively. The fourth column shows the local-standard-of-rest velocity of the spot, the fifth column its peak flux density. The last column presents the brightness temperatures of each spot.}}
\endlastfoot
II & $-$207.86 $\pm$ 0.10 & $-$192.23 $\pm$ 0.08 & $-$0.2 & 0.15 &  5.1$\times$10$^{7}$ \\
II & $-$208.36 $\pm$ 0.07 & $-$192.01 $\pm$ 0.05 & $-$0.2 & 0.28 &  9.2$\times$10$^{7}$ \\
II & $-$208.63 $\pm$ 0.06 & $-$192.35 $\pm$ 0.05 & $-$0.3 & 0.35 &  1.2$\times$10$^{8}$ \\
II & $-$208.92 $\pm$ 0.06 & $-$192.37 $\pm$ 0.04 & $-$0.3 & 0.41 &  1.4$\times$10$^{8}$ \\
II & $-$209.32 $\pm$ 0.06 & $-$192.36 $\pm$ 0.03 & $-$0.4 & 0.43 &  1.4$\times$10$^{8}$ \\
II & $-$208.91 $\pm$ 0.20 & $-$191.45 $\pm$ 0.10 & $-$0.4 & 0.37 &  1.2$\times$10$^{8}$ \\
II & $-$209.88 $\pm$ 0.07 & $-$191.60 $\pm$ 0.04 & $-$0.5 & 0.33 &  1.1$\times$10$^{8}$ \\
II & $-$209.93 $\pm$ 0.07 & $-$191.63 $\pm$ 0.04 & $-$0.5 & 0.29 &  9.5$\times$10$^{7}$ \\
II & $-$210.19 $\pm$ 0.09 & $-$191.56 $\pm$ 0.04 & $-$0.5 & 0.21 &  7.0$\times$10$^{7}$ \\
II & $-$210.73 $\pm$ 0.10 & $-$191.34 $\pm$ 0.05 & $-$0.6 & 0.15 &  4.8$\times$10$^{7}$ \\
II & $-$210.76 $\pm$ 0.12 & $-$191.11 $\pm$ 0.06 & $-$0.6 & 0.11 &  3.6$\times$10$^{7}$ \\
II & $-$211.31 $\pm$ 0.21 & $-$191.72 $\pm$ 0.09 & $-$0.7 & 0.07 &  2.2$\times$10$^{7}$ \\
\\
II & $-$195.58 $\pm$ 0.40 & $-$175.68 $\pm$ 0.21 & $-$1.0 & 0.03 &  1.1$\times$10$^{7}$ \\
II & $-$194.95 $\pm$ 0.47 & $-$175.16 $\pm$ 0.20 & $-$1.0 & 0.04 &  1.2$\times$10$^{7}$ \\
II & $-$195.78 $\pm$ 0.36 & $-$174.66 $\pm$ 0.21 & $-$1.1 & 0.06 &  1.8$\times$10$^{7}$ \\
\\
II & $-$210.91 $\pm$ 0.15 & $-$169.01 $\pm$ 0.08 & $-$1.2 & 0.09 &  3.0$\times$10$^{7}$ \\
II & $-$211.75 $\pm$ 0.12 & $-$168.57 $\pm$ 0.06 & $-$1.3 & 0.15 &  5.0$\times$10$^{7}$ \\
II & $-$211.05 $\pm$ 0.10 & $-$168.60 $\pm$ 0.05 & $-$1.3 & 0.19 &  6.1$\times$10$^{7}$ \\
II & $-$212.23 $\pm$ 0.12 & $-$167.98 $\pm$ 0.06 & $-$1.4 & 0.22 &  7.3$\times$10$^{7}$ \\
II & $-$214.14 $\pm$ 0.25 & $-$166.91 $\pm$ 0.10 & $-$1.4 & 0.23 &  7.6$\times$10$^{7}$ \\
\\
I & $-$517.79 $\pm$ 0.05 & $-$69.56 $\pm$ 0.02 & $-$1.5 & 0.87 &  2.9$\times$10$^{8}$ \\
I & $-$517.89 $\pm$ 0.03 & $-$68.99 $\pm$ 0.01 & $-$1.5 & 1.41 &  4.7$\times$10$^{8}$ \\
I & $-$517.83 $\pm$ 0.02 & $-$68.50 $\pm$ 0.01 & $-$1.6 & 2.16 &  7.1$\times$10$^{8}$ \\
I & $-$518.03 $\pm$ 0.03 & $-$66.28 $\pm$ 0.02 & $-$1.6 & 5.23 &  1.7$\times$10$^{9}$ \\
I & $-$517.76 $\pm$ 0.01 & $-$65.30 $\pm$ 0.01 & $-$1.6 & 8.29 &  2.7$\times$10$^{9}$ \\
I & $-$517.80 $\pm$ 0.01 & $-$65.81 $\pm$ 0.01 & $-$1.7 & 12.30 &  4.1$\times$10$^{9}$ \\
I & $-$518.33 $\pm$ 0.02 & $-$65.79 $\pm$ 0.01 & $-$1.7 & 13.07 &  4.3$\times$10$^{9}$ \\
I & $-$518.44 $\pm$ 0.02 & $-$65.47 $\pm$ 0.01 & $-$1.8 & 15.41 &  5.1$\times$10$^{9}$ \\
I & $-$518.76 $\pm$ 0.01 & $-$65.75 $\pm$ 0.01 & $-$1.8 & 13.99 &  4.6$\times$10$^{9}$ \\
I & $-$519.00 $\pm$ 0.02 & $-$65.84 $\pm$ 0.01 & $-$1.9 & 9.05 &  3.0$\times$10$^{9}$ \\
I & $-$518.69 $\pm$ 0.02 & $-$65.82 $\pm$ 0.01 & $-$1.9 & 4.72 &  1.6$\times$10$^{9}$ \\
I & $-$518.53 $\pm$ 0.05 & $-$65.82 $\pm$ 0.02 & $-$1.9 & 1.41 &  4.7$\times$10$^{8}$ \\
\\
I & $-$587.29 $\pm$ 0.03 & $-$17.73 $\pm$ 0.01 & $-$2.2 & 6.78 &  2.2$\times$10$^{9}$ \\
I & $-$587.76 $\pm$ 0.02 & $-$16.95 $\pm$ 0.01 & $-$2.2 & 23.42 &  7.8$\times$10$^{9}$ \\
I & $-$589.13 $\pm$ 0.02 & $-$16.39 $\pm$ 0.01 & $-$2.3 & 46.93 &  1.6$\times$10$^{10}$ \\
I & $-$590.42 $\pm$ 0.03 & $-$15.96 $\pm$ 0.01 & $-$2.3 & 71.28 &  2.4$\times$10$^{10}$ \\
I & $-$591.24 $\pm$ 0.03 & $-$15.39 $\pm$ 0.01 & $-$2.3 & 69.22 &  2.3$\times$10$^{10}$ \\
I & $-$592.50 $\pm$ 0.03 & $-$14.46 $\pm$ 0.01 & $-$2.4 & 56.08 &  1.9$\times$10$^{10}$ \\
I & $-$594.28 $\pm$ 0.04 & $-$13.29 $\pm$ 0.02 & $-$2.4 & 32.95 &  1.1$\times$10$^{10}$ \\
I & $-$596.01 $\pm$ 0.05 & $-$12.76 $\pm$ 0.02 & $-$2.5 & 24.56 &  8.1$\times$10$^{9}$ \\
I & $-$599.76 $\pm$ 0.09 & $-$11.26 $\pm$ 0.04 & $-$2.5 & 13.67 &  4.5$\times$10$^{9}$ \\
I & $-$602.11 $\pm$ 0.07 & $-$10.22 $\pm$ 0.04 & $-$2.6 & 10.24 &  3.4$\times$10$^{9}$ \\
I & $-$602.80 $\pm$ 0.07 & $-$9.79 $\pm$ 0.04 & $-$2.6 & 7.69 &  2.5$\times$10$^{9}$ \\
I & $-$603.32 $\pm$ 0.12 & $-$10.17 $\pm$ 0.05 & $-$2.6 & 3.05 &  1.0$\times$10$^{9}$ \\
\\
I & $-$624.95 $\pm$ 0.08 & $-$2.75 $\pm$ 0.04 & $-$2.6 & 4.50 &  1.5$\times$10$^{9}$ \\
I & $-$625.86 $\pm$ 0.04 & $-$2.14 $\pm$ 0.02 & $-$2.7 & 4.01 &  1.3$\times$10$^{9}$ \\
I & $-$627.00 $\pm$ 0.05 & $-$1.81 $\pm$ 0.02 & $-$2.7 & 2.00 &  6.6$\times$10$^{8}$ \\
I & $-$628.20 $\pm$ 0.03 & $-$1.50 $\pm$ 0.01 & $-$2.8 & 2.17 &  7.2$\times$10$^{8}$ \\
I & $-$628.53 $\pm$ 0.03 & $-$1.15 $\pm$ 0.01 & $-$2.8 & 1.75 &  5.8$\times$10$^{8}$ \\
I & $-$629.03 $\pm$ 0.04 & $-$0.87 $\pm$ 0.02 & $-$2.9 & 0.70 &  2.3$\times$10$^{8}$ \\
I & $-$629.05 $\pm$ 0.06 & $-$0.64 $\pm$ 0.03 & $-$2.9 & 0.32 &  1.1$\times$10$^{8}$ \\
I & $-$629.31 $\pm$ 0.11 & 0.03 $\pm$ 0.05 & $-$3.0 & 0.15 &  4.9$\times$10$^{7}$ \\
\\
III & $-$762.95 $\pm$ 0.14 & 77.27 $\pm$ 0.07 & $-$3.3 & 0.52 &  1.7$\times$10$^{8}$ \\
III & $-$763.00 $\pm$ 0.14 & 76.80 $\pm$ 0.08 & $-$3.4 & 0.83 &  2.8$\times$10$^{8}$ \\
III & $-$762.65 $\pm$ 0.10 & 76.92 $\pm$ 0.05 & $-$3.4 & 0.73 &  2.4$\times$10$^{8}$ \\
III & $-$763.27 $\pm$ 0.32 & 76.71 $\pm$ 0.16 & $-$3.4 & 0.24 &  8.0$\times$10$^{7}$ \\
III & $-$773.07 $\pm$ 0.07 & 80.50 $\pm$ 0.04 & $-$3.5 & 0.50 &  1.7$\times$10$^{8}$ \\
III & $-$773.72 $\pm$ 0.09 & 81.15 $\pm$ 0.06 & $-$3.5 & 0.36 &  1.2$\times$10$^{8}$ \\
\\
III & $-$827.92 $\pm$ 0.04 & 123.60 $\pm$ 0.02 & $-$3.6 & 1.05 &  3.5$\times$10$^{8}$ \\
III & $-$828.07 $\pm$ 0.06 & 123.20 $\pm$ 0.03 & $-$3.6 & 0.71 &  2.4$\times$10$^{8}$ \\
III & $-$827.79 $\pm$ 0.12 & 123.31 $\pm$ 0.06 & $-$3.7 & 0.38 &  1.3$\times$10$^{8}$ \\
\\
IV & 813.92 $\pm$ 0.21 & $-$209.73 $\pm$ 0.08 & $-$3.7 & 0.19 &  6.4$\times$10$^{7}$ \\
IV & 814.37 $\pm$ 0.11 & $-$209.60 $\pm$ 0.04 & $-$3.7 & 0.38 &  1.3$\times$10$^{8}$ \\
IV & 814.40 $\pm$ 0.06 & $-$209.63 $\pm$ 0.03 & $-$3.8 & 0.85 &  2.8$\times$10$^{8}$ \\
IV & 814.16 $\pm$ 0.05 & $-$209.65 $\pm$ 0.02 & $-$3.8 & 1.16 &  3.8$\times$10$^{8}$ \\
IV & 812.40 $\pm$ 0.19 & $-$209.93 $\pm$ 0.08 & $-$3.9 & 0.37 &  1.2$\times$10$^{8}$ \\
\\
IV & 797.92 $\pm$ 0.03 & $-$209.07 $\pm$ 0.01 & $-$3.8 & 2.19 &  7.2$\times$10$^{8}$ \\
IV & 796.50 $\pm$ 0.02 & $-$209.10 $\pm$ 0.01 & $-$3.9 & 3.33 &  1.1$\times$10$^{9}$ \\
IV & 795.63 $\pm$ 0.03 & $-$209.32 $\pm$ 0.01 & $-$3.9 & 3.15 &  1.0$\times$10$^{9}$ \\
IV & 795.38 $\pm$ 0.05 & $-$209.77 $\pm$ 0.02 & $-$4.0 & 3.74 &  1.2$\times$10$^{9}$ \\
IV & 794.32 $\pm$ 0.05 & $-$209.80 $\pm$ 0.02 & $-$4.0 & 3.63 &  1.2$\times$10$^{9}$ \\
IV & 794.30 $\pm$ 0.04 & $-$209.85 $\pm$ 0.02 & $-$4.1 & 4.18 &  1.4$\times$10$^{9}$ \\
IV & 793.96 $\pm$ 0.05 & $-$209.85 $\pm$ 0.02 & $-$4.1 & 4.00 &  1.3$\times$10$^{9}$ \\
IV & 793.71 $\pm$ 0.05 & $-$209.66 $\pm$ 0.02 & $-$4.1 & 3.92 &  1.3$\times$10$^{9}$ \\
IV & 793.80 $\pm$ 0.04 & $-$209.36 $\pm$ 0.02 & $-$4.2 & 4.12 &  1.4$\times$10$^{9}$ \\
IV & 794.41 $\pm$ 0.03 & $-$208.86 $\pm$ 0.01 & $-$4.3 & 1.98 &  6.6$\times$10$^{8}$ \\
IV & 794.80 $\pm$ 0.04 & $-$208.91 $\pm$ 0.02 & $-$4.3 & 1.11 &  3.7$\times$10$^{8}$ \\
IV & 795.05 $\pm$ 0.09 & $-$208.78 $\pm$ 0.04 & $-$4.4 & 0.38 &  1.3$\times$10$^{8}$ \\
IV & 795.13 $\pm$ 0.13 & $-$208.83 $\pm$ 0.07 & $-$4.4 & 0.13 &  4.3$\times$10$^{7}$ \\
\\
IV & 796.75 $\pm$ 0.07 & $-$224.11 $\pm$ 0.03 & $-$3.9 & 1.46 &  4.8$\times$10$^{8}$ \\
IV & 796.40 $\pm$ 0.04 & $-$224.12 $\pm$ 0.02 & $-$4.0 & 5.21 &  1.7$\times$10$^{9}$ \\
IV & 797.03 $\pm$ 0.03 & $-$223.85 $\pm$ 0.01 & $-$4.0 & 6.09 &  2.0$\times$10$^{9}$ \\
IV & 797.08 $\pm$ 0.03 & $-$223.59 $\pm$ 0.01 & $-$4.1 & 5.84 &  1.9$\times$10$^{9}$ \\
IV & 796.71 $\pm$ 0.05 & $-$223.70 $\pm$ 0.02 & $-$4.1 & 3.80 &  1.3$\times$10$^{9}$ \\
IV & 796.58 $\pm$ 0.06 & $-$223.63 $\pm$ 0.03 & $-$4.1 & 3.09 &  1.0$\times$10$^{9}$ \\
IV & 796.23 $\pm$ 0.10 & $-$223.64 $\pm$ 0.04 & $-$4.2 & 1.49 &  4.9$\times$10$^{8}$ \\
\\
V & $-$482.68 $\pm$ 0.39 & 548.40 $\pm$ 0.41 & $-$4.3 & 0.14 &  4.7$\times$10$^{7}$ \\
V & $-$437.63 $\pm$ 0.28 & 596.95 $\pm$ 0.17 & $-$4.3 & 0.18 &  5.8$\times$10$^{7}$ \\
V & $-$448.15 $\pm$ 0.27 & 583.80 $\pm$ 0.32 & $-$4.4 & 0.13 &  4.4$\times$10$^{7}$ \\
V & $-$482.35 $\pm$ 0.28 & 547.95 $\pm$ 0.32 & $-$4.4 & 0.13 &  4.2$\times$10$^{7}$ \\
\end{longtable}

\newpage
\begin{longtable}{p{0.15\textwidth}p{0.15\textwidth}p{0.15\textwidth}p{0.15\textwidth}p{0.15\textwidth}p{0.15\textwidth}}
\caption{
List of 6.7 GHz CH$_3$OH maser spots detected in epoch 9.
\label{tab:spots_ed048B} }
\\\hline
\noalign{\vskip 1mm}  
Cluster & $\Delta$\,$\alpha$ & $\Delta$\,$\delta$ & V$_{\mathrm{lsr}}$ & F$_{\mathrm{peak}}$  & T$_{\mathrm{b}}$ \\
 & (mas) & (mas) & (\kms) & (Jy\,beam$^{-1}$) & (K)
\\\hline
\noalign{\vskip 1mm}  
\endfirsthead
\caption{continued}\\
\\\hline
\noalign{\vskip 1mm}  
Cluster & $\Delta$\,$\alpha$ & $\Delta$\,$\delta$ & V$_{\mathrm{lsr}}$ & F$_{\mathrm{peak}}$  & T$_{\mathrm{b}}$ \\
 & (mas) & (mas) & (\kms) & (Jy\,beam$^{-1}$) & (K)
\\\hline
\noalign{\vskip 1mm}  
\endhead
\hline
\endfoot
\hline
\noalign{\vskip 1mm}
\multicolumn{6}{r}{\tablefoot{Numbers in the first column refer to the cluster that the spot originates from (see figure \ref{fig:cepa-spots-june-and-october-2020}). Columns 2 and 3 give the relative positions and uncertainties in Right Ascension and Declination, respectively. The fourth column shows the local-standard-of-rest velocity of the spot, and the fifth column its peak flux density. The last column presents the brightness temperatures of each spot.}}
\endlastfoot
II & $-$208.33 $\pm$ 0.07 & $-$191.05 $\pm$ 0.10 & $-$0.1 & 0.13 &  6.7$\times$10$^{7}$ \\
II & $-$208.23 $\pm$ 0.16 & $-$190.07 $\pm$ 0.24 & $-$0.1 & 0.06 &  2.9$\times$10$^{7}$ \\
II & $-$208.39 $\pm$ 0.04 & $-$191.97 $\pm$ 0.06 & $-$0.2 & 0.23 &  1.1$\times$10$^{8}$ \\
II & $-$208.60 $\pm$ 0.03 & $-$192.39 $\pm$ 0.04 & $-$0.3 & 0.29 &  1.4$\times$10$^{8}$ \\
II & $-$209.13 $\pm$ 0.04 & $-$192.15 $\pm$ 0.04 & $-$0.4 & 0.26 &  1.3$\times$10$^{8}$ \\
II & $-$209.83 $\pm$ 0.04 & $-$191.80 $\pm$ 0.05 & $-$0.5 & 0.19 &  9.6$\times$10$^{7}$ \\
II & $-$210.57 $\pm$ 0.06 & $-$191.27 $\pm$ 0.07 & $-$0.6 & 0.12 &  6.1$\times$10$^{7}$ \\
II & $-$211.20 $\pm$ 0.10 & $-$191.00 $\pm$ 0.12 & $-$0.7 & 0.06 &  2.9$\times$10$^{7}$ \\
\\
I & $-$517.76 $\pm$ 0.02 & $-$68.79 $\pm$ 0.01 & $-$1.5 & 1.91 &  9.6$\times$10$^{8}$ \\
I & $-$517.66 $\pm$ 0.02 & $-$67.43 $\pm$ 0.01 & $-$1.6 & 6.72 &  3.4$\times$10$^{9}$ \\
I & $-$517.65 $\pm$ 0.01 & $-$66.06 $\pm$ 0.01 & $-$1.6 & 20.20 &  1.0$\times$10$^{10}$ \\
I & $-$517.80 $\pm$ 0.01 & $-$65.81 $\pm$ 0.01 & $-$1.7 & 35.25 &  1.8$\times$10$^{10}$ \\
I & $-$518.04 $\pm$ 0.02 & $-$65.83 $\pm$ 0.01 & $-$1.8 & 29.48 &  1.5$\times$10$^{10}$ \\
I & $-$518.24 $\pm$ 0.02 & $-$65.89 $\pm$ 0.01 & $-$1.9 & 11.80 &  5.9$\times$10$^{9}$ \\
I & $-$518.18 $\pm$ 0.05 & $-$65.93 $\pm$ 0.03 & $-$2.0 & 2.17 &  1.1$\times$10$^{9}$ \\
\\
I & $-$556.77 $\pm$ 0.04 & $-$34.17 $\pm$ 0.04 & $-$2.1 & 3.84 &  1.9$\times$10$^{9}$ \\
I & $-$556.63 $\pm$ 0.09 & $-$33.47 $\pm$ 0.09 & $-$2.2 & 4.44 &  2.2$\times$10$^{9}$ \\
\\
I & $-$572.04 $\pm$ 0.05 & $-$49.10 $\pm$ 0.03 & $-$2.0 & 2.83 &  1.4$\times$10$^{9}$ \\
I & $-$573.15 $\pm$ 0.07 & $-$47.27 $\pm$ 0.04 & $-$2.1 & 3.42 &  1.7$\times$10$^{9}$ \\
\\
I & $-$587.18 $\pm$ 0.03 & $-$17.36 $\pm$ 0.02 & $-$2.2 & 23.16 &  1.2$\times$10$^{10}$ \\
I & $-$589.01 $\pm$ 0.02 & $-$16.28 $\pm$ 0.01 & $-$2.3 & 76.96 &  3.9$\times$10$^{10}$ \\
I & $-$590.81 $\pm$ 0.02 & $-$15.21 $\pm$ 0.02 & $-$2.3 & 112.29 &  5.6$\times$10$^{10}$ \\
I & $-$593.35 $\pm$ 0.03 & $-$13.92 $\pm$ 0.02 & $-$2.4 & 86.62 &  4.4$\times$10$^{10}$ \\
I & $-$598.13 $\pm$ 0.05 & $-$11.75 $\pm$ 0.02 & $-$2.5 & 38.25 &  1.9$\times$10$^{10}$ \\
I & $-$602.13 $\pm$ 0.04 & $-$9.95 $\pm$ 0.02 & $-$2.6 & 14.02 &  7.0$\times$10$^{9}$ \\
I & $-$603.01 $\pm$ 0.09 & $-$9.28 $\pm$ 0.06 & $-$2.7 & 2.49 &  1.3$\times$10$^{9}$ \\
\\
I & $-$619.39 $\pm$ 0.09 & $-$3.87 $\pm$ 0.04 & $-$2.6 & 7.15 &  3.6$\times$10$^{9}$ \\
I & $-$623.67 $\pm$ 0.04 & $-$2.46 $\pm$ 0.02 & $-$2.7 & 7.57 &  3.8$\times$10$^{9}$ \\
I & $-$626.71 $\pm$ 0.03 & $-$1.26 $\pm$ 0.02 & $-$2.8 & 5.71 &  2.9$\times$10$^{9}$ \\
I & $-$628.29 $\pm$ 0.02 & $-$0.28 $\pm$ 0.02 & $-$2.9 & 2.49 &  1.3$\times$10$^{9}$ \\
I & $-$629.04 $\pm$ 0.04 & 0.53 $\pm$ 0.03 & $-$3.0 & 0.59 &  3.0$\times$10$^{8}$ \\
I & $-$629.45 $\pm$ 0.24 & 1.43 $\pm$ 0.12 & $-$3.0 & 0.08 &  3.9$\times$10$^{7}$ \\
\\
III & $-$773.27 $\pm$ 0.03 & 80.12 $\pm$ 0.04 & $-$3.5 & 1.43 &  7.2$\times$10$^{8}$ \\
III & $-$773.16 $\pm$ 0.03 & 80.00 $\pm$ 0.03 & $-$3.5 & 1.45 &  7.3$\times$10$^{8}$ \\
III & $-$773.27 $\pm$ 0.05 & 79.07 $\pm$ 0.07 & $-$3.6 & 1.11 &  5.6$\times$10$^{8}$ \\
\\
III & $-$768.41 $\pm$ 0.07 & 85.20 $\pm$ 0.08 & $-$3.1 & 0.13 &  6.7$\times$10$^{7}$ \\
III & $-$768.00 $\pm$ 0.03 & 83.58 $\pm$ 0.03 & $-$3.2 & 0.61 &  3.0$\times$10$^{8}$ \\
III & $-$767.36 $\pm$ 0.04 & 81.46 $\pm$ 0.04 & $-$3.3 & 0.85 &  4.3$\times$10$^{8}$ \\
III & $-$769.70 $\pm$ 0.13 & 80.52 $\pm$ 0.06 & $-$3.4 & 0.69 &  3.5$\times$10$^{8}$ \\
\\
III & $-$761.78 $\pm$ 0.03 & 76.31 $\pm$ 0.04 & $-$3.5 & 1.11 &  5.6$\times$10$^{8}$ \\
III & $-$762.37 $\pm$ 0.05 & 78.85 $\pm$ 0.06 & $-$3.2 & 0.38 &  1.9$\times$10$^{8}$ \\
III & $-$762.12 $\pm$ 0.02 & 77.39 $\pm$ 0.02 & $-$3.3 & 1.60 &  8.0$\times$10$^{8}$ \\
III & $-$762.10 $\pm$ 0.02 & 76.73 $\pm$ 0.02 & $-$3.4 & 2.18 &  1.1$\times$10$^{9}$ \\
III & $-$761.78 $\pm$ 0.03 & 76.32 $\pm$ 0.04 & $-$3.5 & 1.11 &  5.6$\times$10$^{8}$ \\
\\
III & $-$827.10 $\pm$ 0.07 & 124.00 $\pm$ 0.05 & $-$3.5 & 0.79 &  4.0$\times$10$^{8}$ \\
III & $-$827.34 $\pm$ 0.04 & 123.36 $\pm$ 0.02 & $-$3.6 & 2.40 &  1.2$\times$10$^{9}$ \\
III & $-$827.50 $\pm$ 0.04 & 122.77 $\pm$ 0.02 & $-$3.7 & 2.52 &  1.3$\times$10$^{9}$ \\
III & $-$828.75 $\pm$ 0.09 & 121.56 $\pm$ 0.05 & $-$3.7 & 1.14 &  5.7$\times$10$^{8}$ \\
\\
III & $-$870.97 $\pm$ 0.04 & 196.99 $\pm$ 0.05 & $-$3.6 & 1.40 &  7.0$\times$10$^{8}$ \\
III & $-$871.86 $\pm$ 0.04 & 197.33 $\pm$ 0.04 & $-$3.7 & 1.88 &  9.4$\times$10$^{8}$ \\
III & $-$873.28 $\pm$ 0.07 & 198.00 $\pm$ 0.07 & $-$3.7 & 0.91 &  4.6$\times$10$^{8}$ \\
\\
IV & 813.92 $\pm$ 0.06 & $-$209.04 $\pm$ 0.03 & $-$3.7 & 1.34 &  6.7$\times$10$^{8}$ \\
IV & 814.42 $\pm$ 0.08 & $-$208.86 $\pm$ 0.04 & $-$3.8 & 1.78 &  9.0$\times$10$^{8}$ \\
\\
IV & 799.44 $\pm$ 0.04 & $-$209.69 $\pm$ 0.03 & $-$3.7 & 1.26 &  6.3$\times$10$^{8}$ \\
IV & 798.33 $\pm$ 0.02 & $-$209.50 $\pm$ 0.01 & $-$3.8 & 6.58 &  3.3$\times$10$^{9}$ \\
IV & 797.07 $\pm$ 0.02 & $-$209.48 $\pm$ 0.01 & $-$3.9 & 14.58 &  7.3$\times$10$^{9}$ \\
IV & 796.05 $\pm$ 0.02 & $-$209.56 $\pm$ 0.01 & $-$4.0 & 19.96 &  1.0$\times$10$^{10}$ \\
IV & 795.51 $\pm$ 0.02 & $-$209.61 $\pm$ 0.01 & $-$4.1 & 19.20 &  9.6$\times$10$^{9}$ \\
IV & 795.41 $\pm$ 0.02 & $-$209.50 $\pm$ 0.01 & $-$4.2 & 13.44 &  6.8$\times$10$^{9}$ \\
IV & 795.44 $\pm$ 0.01 & $-$209.32 $\pm$ 0.01 & $-$4.3 & 5.98 &  3.0$\times$10$^{9}$ \\
IV & 795.34 $\pm$ 0.02 & $-$209.07 $\pm$ 0.02 & $-$4.4 & 1.62 &  8.1$\times$10$^{8}$ \\
IV & 795.35 $\pm$ 0.05 & $-$208.95 $\pm$ 0.04 & $-$4.4 & 0.27 &  1.4$\times$10$^{8}$ \\
\\
IV & 796.09 $\pm$ 0.10 & $-$224.76 $\pm$ 0.06 & $-$3.8 & 1.49 &  7.5$\times$10$^{8}$ \\
IV & 795.66 $\pm$ 0.03 & $-$224.71 $\pm$ 0.02 & $-$3.9 & 11.12 &  5.6$\times$10$^{9}$ \\
IV & 795.72 $\pm$ 0.03 & $-$224.53 $\pm$ 0.01 & $-$4.0 & 21.93 &  1.1$\times$10$^{10}$ \\
IV & 795.70 $\pm$ 0.03 & $-$224.30 $\pm$ 0.01 & $-$4.1 & 18.53 &  9.3$\times$10$^{9}$ \\
IV & 795.14 $\pm$ 0.04 & $-$224.10 $\pm$ 0.02 & $-$4.2 & 8.82 &  4.4$\times$10$^{9}$ \\
IV & 794.21 $\pm$ 0.06 & $-$224.01 $\pm$ 0.03 & $-$4.3 & 2.23 &  1.1$\times$10$^{9}$ \\
\\
V & $-$423.83 $\pm$ 0.10 & 485.40 $\pm$ 0.09 & $-$4.6 & 0.11 &  5.6$\times$10$^{7}$ \\
V & $-$425.20 $\pm$ 0.09 & 486.14 $\pm$ 0.09 & $-$4.7 & 0.12 &  6.3$\times$10$^{7}$ \\
V & $-$426.98 $\pm$ 0.19 & 487.12 $\pm$ 0.19 & $-$4.8 & 0.06 &  3.0$\times$10$^{7}$ \\
\end{longtable}

\begin{table}
\centering
\caption{Cloudlet parameters derived from Gaussian function fitting. \label{tab:cloudlet_gauss_fit_results} }
\begin{tabular}[c]{lccc}
\hline
Epoch & $F_{\mathrm{p}}$ & $V_{\mathrm{p}}$ & FWHM \\
      & (Jy\,beam$^{-1}$)   & (km\,s$^{-1}$)      & (km\,s$^{-1}$) \\
\hline
\noalign{\vskip 1mm}  
\multicolumn{4}{l}{\bf $-$0.5\,km\,s$^{-1}$ cloudlet}\\
7 & 0.89 $\pm$ 0.02 & $-$0.462 $\pm$ 0.003 & 0.33 $\pm$ 0.01 \\
8 & 0.42 $\pm$ 0.02 & $-$0.377 $\pm$ 0.005 & 0.34 $\pm$ 0.01 \\
9 & 0.28 $\pm$ 0.01 & $-$0.355 $\pm$ 0.004 & 0.40 $\pm$ 0.01 \\
\multicolumn{4}{l}{\bf $-$1.3\,km\,s$^{-1}$ cloudlet}\\
7 & 0.24 $\pm$ 0.01 & $-$1.177 $\pm$ 0.002 & 0.18 $\pm$ 0.01 \\
8 & 0.23 $\pm$ 0.01 & $-$1.405 $\pm$ 0.022 & 0.28 $\pm$ 0.04 \\
\multicolumn{4}{l}{\bf $-$1.7\,km\,s$^{-1}$ cloudlet}\\
5 & 11.85 $\pm$ 0.11 & $-$1.740 $\pm$ 0.001 & 0.27 $\pm$ 0.01 \\
6 & 23.17 $\pm$ 0.69 & $-$1.746 $\pm$ 0.002 & 0.20 $\pm$ 0.01 \\
7 & 36.82 $\pm$ 1.41 & $-$1.745 $\pm$ 0.002 & 0.21 $\pm$ 0.01 \\
8 & 14.98 $\pm$ 0.99 & $-$1.746 $\pm$ 0.005 & 0.24 $\pm$ 0.01 \\
9 & 32.73 $\pm$ 2.99 & $-$1.742 $\pm$ 0.006 & 0.26 $\pm$ 0.01 \\
\multicolumn{4}{l}{\bf $-$2.6\,km\,s$^{-1}$ cloudlet}\\
5 &  19.19 $\pm$ 0.28   & $-$2.491 $\pm$ 0.002 & 0.30 $\pm$ 0.01 \\
6 &  45.51 $\pm$ 0.83   & $-$2.460 $\pm$ 0.002 & 0.24 $\pm$ 0.01 \\
7 & 116.32 $\pm$ 1.83   & $-$2.443 $\pm$ 0.001 & 0.23 $\pm$ 0.01 \\
8 &  67.48 $\pm$ 16.49  & $-$2.317 $\pm$ 0.009 & 0.17 $\pm$ 0.01 \\
  &  16.43 $\pm$ 7.98   & $-$2.460 $\pm$ 0.097 & 0.25 $\pm$ 0.10 \\
9 &  97.50 $\pm$ 84.55  & $-$2.326 $\pm$ 0.055 & 0.22 $\pm$ 0.04 \\
  &  32.34 $\pm$ 72.41  & $-$2.460 $\pm$ 0.222 & 0.25 $\pm$ 0.13 \\
\multicolumn{4}{l}{\bf $-$2.7\,km\,s$^{-1}$ cloudlet}\\
5 & 6.95 $\pm$ 0.05 & $-$2.739 $\pm$ 0.002 & 0.28 $\pm$ 0.01 \\
6 & 6.95 $\pm$ 0.06 & $-$2.729 $\pm$ 0.002 & 0.25 $\pm$ 0.01 \\
7 & 9.23 $\pm$ 0.08 & $-$2.717 $\pm$ 0.002 & 0.25 $\pm$ 0.01 \\
8 & 3.63 $\pm$ 0.96 & $-$2.663 $\pm$ 0.059 & 0.27 $\pm$ 0.06 \\
9 & 7.93 $\pm$ 0.29 & $-$2.684 $\pm$ 0.008 & 0.28 $\pm$ 0.01 \\
\multicolumn{4}{l}{\bf $-$3.4\,km\,s$^{-1}$ cloudlet}\\
5 & 1.16 $\pm$ 0.01 & $-$3.430 $\pm$ 0.004 & 0.31 $\pm$ 0.01 \\
6 & 1.93 $\pm$ 0.07 & $-$3.426 $\pm$ 0.003 & 0.18 $\pm$ 0.01 \\
7 & 2.87 $\pm$ 0.09 & $-$3.413 $\pm$ 0.003 & 0.17 $\pm$ 0.01 \\
8 & 0.90 $\pm$ 0.06 & $-$3.361 $\pm$ 0.003 & 0.12 $\pm$ 0.01 \\
9 & 2.24 $\pm$ 0.03 & $-$3.381 $\pm$ 0.001 & 0.20 $\pm$ 0.00 \\
\multicolumn{4}{l}{\bf $-$3.6\,km\,s$^{-1}$ cloudlet}\\
5 & 5.32 $\pm$ 0.07 & $-$3.676 $\pm$ 0.002 & 0.27 $\pm$ 0.01 \\
6 & 5.16 $\pm$ 0.27 & $-$3.667 $\pm$ 0.004 & 0.19 $\pm$ 0.01 \\
7 & 5.20 $\pm$ 0.18 & $-$3.646 $\pm$ 0.003 & 0.20 $\pm$ 0.01 \\
8 & 1.22 $\pm$ ---  & $-$3.521 $\pm$ ---   & 0.21 $\pm$ ---  \\
9 & 2.79 $\pm$ 0.10 & $-$3.625 $\pm$ 0.003 & 0.21 $\pm$ 0.01 \\
\multicolumn{4}{l}{\bf $-$4.0\,km\,s$^{-1}$ cloudlet \#1}\\
5 & 12.87 $\pm$ 0.09 & $-$4.089 $\pm$ 0.001 & 0.36 $\pm$ 0.01 \\
6 & 11.71 $\pm$ 0.48 & $-$4.078 $\pm$ 0.004 & 0.33 $\pm$ 0.01 \\
7 & 15.06 $\pm$ 0.47 & $-$4.097 $\pm$ 0.002 & 0.28 $\pm$ 0.01 \\
8 &  4.77 $\pm$ 0.48 & $-$4.034 $\pm$ 0.011 & 0.37 $\pm$ 0.02 \\
9 & 22.02 $\pm$ 1.01 & $-$4.052 $\pm$ 0.003 & 0.32 $\pm$ 0.01 \\
\multicolumn{4}{l}{\bf $-$4.0\,km\,s$^{-1}$ cloudlet \#2}\\
5 & 8.47 $\pm$ 0.57 & $-$4.203 $\pm$ 0.011 & 0.25 $\pm$ 0.01 \\
  & 5.91 $\pm$ 0.54 & $-$4.003 $\pm$ 0.012 & 0.22 $\pm$ 0.02 \\
6 & 9.27 $\pm$ 0.06 & $-$4.190 $\pm$ 0.001 & 0.17 $\pm$ 0.01 \\
  & 11.06 $\pm$ 0.04 & $-$4.022 $\pm$ 0.001 & 0.16 $\pm$ 0.01 \\
7 & 25.80 $\pm$ 1.40 & $-$4.060 $\pm$ 0.004 & 0.22 $\pm$ 0.01 \\
8 & 6.23 $\pm$ 0.69 & $-$4.040 $\pm$ 0.009 & 0.18 $\pm$ 0.02 \\
9 & 22.55 $\pm$ 1.78 & $-$4.051 $\pm$ 0.006 & 0.23 $\pm$ 0.01 \\
\multicolumn{4}{l}{\bf $-$4.7\,km\,s$^{-1}$ cloudlet}\\
5 & 0.95 $\pm$ 0.01 & $-$4.711 $\pm$ 0.002 & 0.27 $\pm$ 0.01 \\
6 & 0.57 $\pm$ 0.01 & $-$4.716 $\pm$ 0.002 & 0.17 $\pm$ 0.01 \\
7 & 0.42 $\pm$ 0.01 & $-$4.715 $\pm$ 0.003 & 0.18 $\pm$ 0.01 \\
9 & 0.13 $\pm$ ---  & $-$4.680 $\pm$ ---   & 0.22 $\pm$ ---  \\
\hline
\end{tabular}
\tablefoot{The first column presents epochs (listed in Table \ref{tab:evn_projs}). Columns 2\,$-$\,4 show parameters derived from fitting the peak flux density, the peak velocity relative to the local standard of rest, and the full width at half maximum. The uncertainties provided are derived from the covariance matrix.}
\end{table}

\FloatBarrier
\section{Cluster II maps across 9 epochs}
\begin{figure}[h]
\centering
\includegraphics[width=0.45\textwidth]{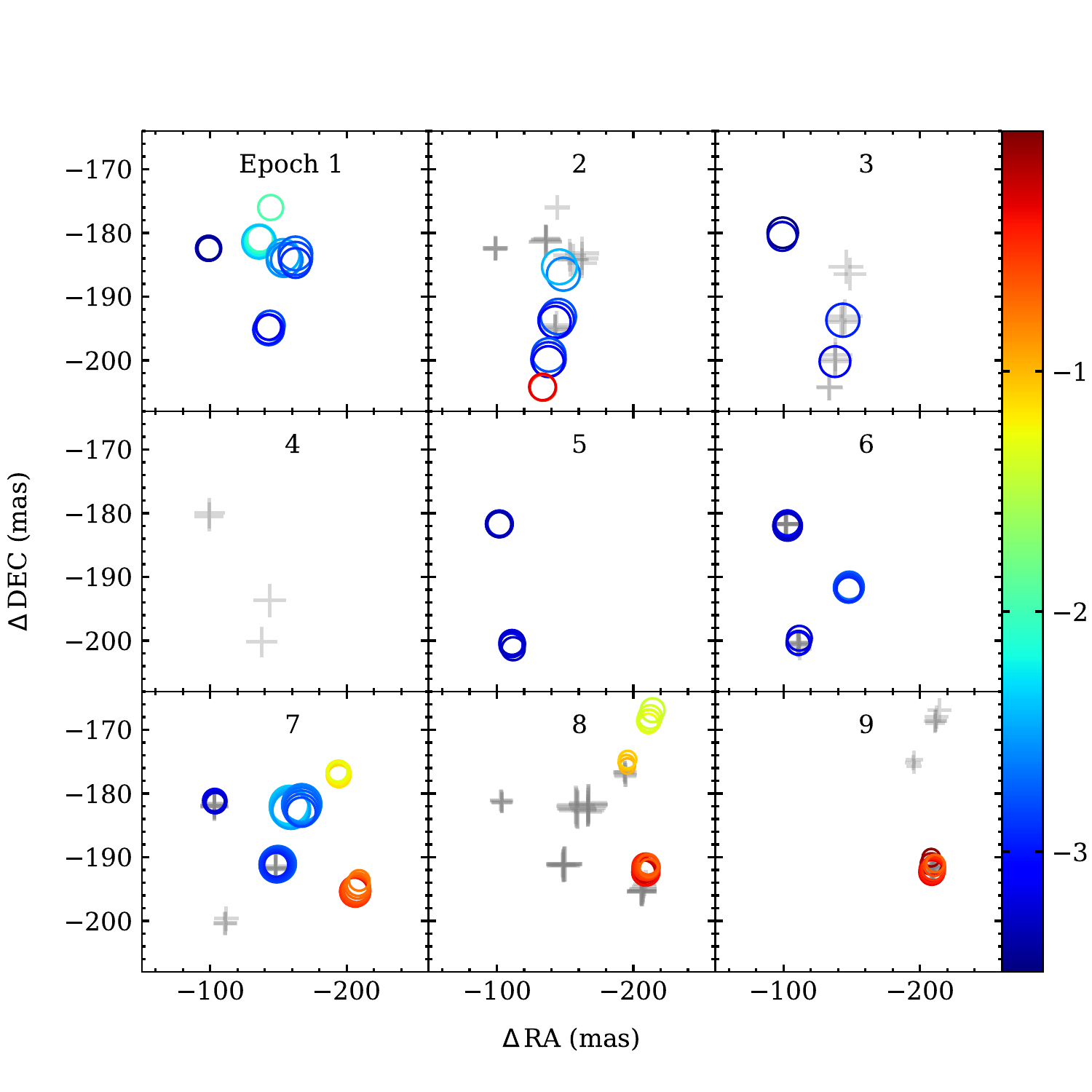}
\caption
{ 
    Spatial distribution of the CH$_3$OH maser spots in Cep\,A\,HW2 in cluster II across nine epochs. Epoch numbers presented at the top of the graphs are listed in Table \ref{tab:evn_projs}. The sizes of the symbols are proportional to the logarithm of the spot peak flux densities. Grey crosses denote the spatial distribution of the emission in the previous epoch. \label{fig:cluster_I_and_II_8_epochs} 
}
\end{figure}
\end{appendix}
\end{document}